\documentclass[a4paper,fleqn,usenatbib]{mnras}
\usepackage{newtxtext,newtxmath}
 \usepackage{xcolor}
\usepackage[T1]{fontenc}
\usepackage{ae,aecompl}

\usepackage{graphicx}	
\usepackage{amsmath}	

\title[S2CLS/EGS III: faint SMGs and SFGs]
{The SCUBA-2 Cosmology Legacy Survey: The EGS deep field - III. The evolution of faint submillimeter galaxies at $z<4$}

\author[L. Cardona Torres]{L. Cardona-Torres,$^{1}$\thanks{E-mail: lucardona@inaoep.mx}
I. Aretxaga,$^{1}$ A. Monta\~na,$^{1}$ J. A. Zavala$^{2,3}$ and S.M. Faber$^{4}$
\\
$^{1}$ Instituto Nacional de Astrof\'isica, \'Optica y Electr\'onica (INAOE), Luis Enrique Erro 1, Sta. Ma. Tonantzintla, 72840, Puebla, Mexico\\
$^{2}$ Department of Astronomy, The University of Texas at Austin, 2515 Speedway Blvd Stop C1400, Austin, TX 78712, USA\\
$^{3}$ National Astronomical Observatory of Japan, 2-21-1 Osawa, Mitaka, Tokyo 181-8588, Japan\\
$^{4}$ University of California Observatories/Lick Observatory, University of California, Santa Cruz, CA 95064, USA
}

\date{Accepted XXX. Received YYY; in original form ZZZ}
\pubyear{2022}

\begin{document}
\label{firstpage}
\pagerange{\pageref{firstpage}--\pageref{lastpage}}
\maketitle

\begin{abstract}
    We present a demographic analysis of the physical and morphological properties of $450/850~\mu\rm m$-selected galaxies from the deep observations of the SCUBA-2 Cosmology Legacy Survey in the Extended Groth Strip that are detected below the classical submillimeter-galaxy regime ($S_{850 \mu\rm m}\lesssim 6~\rm mJy$/beam) and compare them with a sample of optically-selected star-forming galaxies detected in the Cosmic Assembly Near-infrared Deep Extragalactic Legacy Survey in the same field. We derive the evolution of the main sequence of star-forming galaxies,
    finding a steeper specific star formation rate versus stellar mass at $z>2.5$ than previous studies. Most faint submillimeter-galaxies fall within $3\sigma$ of the main sequence, but 40~per cent are classified as starbursts. Faint submillimeter galaxies have 50~per cent larger sizes at $2<z<3$ than optically-selected star-forming galaxies of the same mass range.
    This is also the redshift bin where we find the largest fraction of starbursts, and hence we could be witnessing merging processes, as confirmed by the preference for visual-morphology classifications of these systems as irregular disk galaxies and mergers. Both populations show an increment towards lower redshifts ($z<2$) of their concentration in $H$-band morphology, but faint submillimeter galaxies on average show larger concentration values at later times. These findings support the claim that faint submillimeter galaxies are mostly a population of massive dust-obscured disk-like galaxies that develop larger bulge components at later epochs.
    While the similarities are great, the median sizes, starburst numbers and $H$-band concentration of faint submillimeter galaxies differ from those of optically-selected star-forming galaxies of the same stellar mass.
\end{abstract}



\section{Introduction}

The initial studies of the Cosmic Infrared Background with the Cosmic Background Explorer \citep[COBE,][]{Puget_COBE,COBE_implications} led to the discovery of a high redshift population of galaxies with strong far infrared (FIR) emission. These galaxies were first detected at $850~\mu\rm m$ with the James Clerk Maxwell Telescope and were named  submillimeter galaxies \citep[SMGs;][]{Smail1997, Hughes1998, Barger_SMGs1998}. The discovery indicated that there was a considerable amount of stellar emission obscured by dust in the high redshift Universe. SMGs are typically located at high redshifts ($z>1$), have large infrared luminosities ($L_{\rm IR} \gtrsim 10^{12}~\rm L_{\odot}$), star formation rates (SFRs$\gtrsim300 ~\rm{M_{\odot}~yr^{-1}}$) and high gas reservoirs ($10^{10-11}~\rm{M}_{\odot}$) \citep[e.g.][]{Blain2004, Chapman2005, Aretxaga2007_photz, MinYun2012, Casey2014,DaCunha_2015,Cowie2017,Michalowski2017, STUDIES_450-850, Birkin2021,Chen2022_AS2COSPEC}. The early submillimeter surveys traced with single-dish telescopes reached a population of galaxies with flux densities $S_{850 \mu\rm m}\gtrsim 6~\rm mJy$, which we will refer to as classical SMGs \citep{Smail1997, Hughes1998, Coppin_2006,Geach2017,Simpson_2019}.

The counterparts of SMGs were typically found in deep optical, IR and radio imaging by exploiting the radio-submillimeter correlation \citep{Carilli_yun_radio-submm}. This allowed the exploration of the multiwavelength properties of SMGs \citep[i.e][]{Chapman2005, Targett2013, Zavala2018, STUDIES-III_450_Lim2020, STUDIES_450-850}, and their role in the cosmic history of star formation \citep{MadauDickinson_2014}. SMGs were initially characterized as extreme star-forming galaxies. They were associated with major mergers that enhanced the star-formation activity through their interactions \citep[][]{Tacconi_2008}. On the other hand, there is a population of SMGs that lie within the scatter of the high-mass end of the star-formation main sequence, but appear to have enhanced star formation efficiency \citep{Dave_secularSMGs}. The exploration of high-resolution FIR surveys have allowed the detection of two sub-populations of SMGs: starburst and main sequence galaxies with a compact core component and main sequence galaxies with an extended dimmer component \citep[i.e.][]{Simpson_2015, Michalowski2017, Elbaz2018, Gullberg_2019,Tadaki_2020,Puglisi2021}. The compact FIR emission is associated with a post-starburst phase due to the low gas fraction detected in these galaxies that could be explained with a history of strong inflow mergers, which enhance the star formation efficiency \citep{DeckerFrench_2021_PSBs}.

Different studies have used interferometric follow-up of single dish surveys \citep[e.g.][]{Hodge_2013_multiplicity, DaCunha_2015, Miettinen2017_ACosmos_VLA3GHz} to explore SMG propeties at higher resolution. They have shown that there is a $\sim26$~per cent chance of finding multiple counterparts to SMGs with fluxes brighter than $S_{850 \mu \rm m} \geq 5~\rm mJy$  \citep{Stach19}. However, fainter sources suffer to a lesser degree this effect.

New facilities with better resolution and sensitivity allow the exploration of this fainter population of dusty galaxies. For instance, \citet{Ono_2014} studied a sample of 11 dusty star forming galaxies below the SMG regime ($S_{\rm 1.2 mm}\sim 0.1-1.0~\rm mJy$) with ALMA, and found that these are FIR counterparts of UV-selected or K-selected galaxies, like Lyman-break galaxies (LBGs) or star-forming BzK galaxies. \citet{Aravena20} analysed a sample of 32 galaxies detected at 1.2~mm with ALMA as part of the ASPECS Large Program. They estimated a median redshift of $z=1.85$ (with interquartile range $1.10 - 2.57$) and found that 34~per cent of their $S_{1.2\rm{mm}}\sim 0.03 - 1\ \rm{mJy}$ galaxies are located below the main sequence of star formation.

Faint SMGs at higher redshifts have also been detected by means of gravitational lensing magnification effects by either massive galaxies or galaxy clusters in the foreground \citep[e.g.][]{Chen2013_faintlensed450, Aguirre2018_laboca_szecluster}. These, however, are generally limited to small samples biased towards higher redshifts 
and smaller compact objects \cite[e.g.][]{Bussmann2013_lens_SMA_Hatlas-Hermes}. Furthermore, accurate modelling of the lensing masses is required to estimate the intrinsic physical and morphological properties of the lensed sources.

The \textit{H}-band morphology of SMGs has been previously explored for both bright SMGs \citep[LESS/ALESS, $S_{870\mu\rm m} > 3~\rm mJy$; ][]{Targett2013, Chen2015} and $450~\mu\rm m$-selected faint SMGs \citep[STUDIES, $S_{450\mu\rm m}=2.8 - 29.6~\rm mJy$; ][]{Chang2018}. These studies found that SMGs have large rest-frame optical sizes, with disk-like and perturbed morphologies. On the other hand, sub-arcsec angular resolution observations with ALMA of both bright \citep[$S_{870\mu\rm m}= 8 - 16~\rm mJy$; ][]{Simpson_2015} and faint \citep[$S_{1.1\rm mm} > 0.5~\rm mJy$; ][]{Franco20} SMGs have shown that their FIR emission is associated to more compact structures, suggesting spheroid build-up in these galaxies. The early released data from the James Webb Space Telescope (JWST) has confirmed the bulge build-up for a small sample of 7 SMGs in the EGS and UDS fields \citep[][]{Chen2022_JWST_morphSMG}. They studied the morphology by fitting two component S\'ersic models and observed residual spiral arms structures, as well as tidal remnants and clumps. Some of these features were also discovered in the JWST data of a lensed grand design spiral galaxy detected with ALMA  \citep[][]{Wu2022_JWST_ALMA} and 2 SMGs detected in the SMACS J0723.3–7327 cluster \citep{ChengCheng2022_JWST_SMG}.

In paper I of this series \citet{Zavala2017} presented the deepest submillimeter observations of the SCUBA-2 Cosmology Legacy Survey (S2CLS) in the Extended Groth Strip (EGS), providing simultaneous maps at both 450 and $850~\mu\rm m$ and the detection of 144 galaxies with flux densities in the ranges $S_{850 \mu\rm m}=0.7-6~\rm mJy$ and $S_{450 \mu\rm m}=3-17~\rm mJy$, in the flux density regime below classical submillimeter galaxies ($S_{850 \mu\rm m} \lesssim 6 \rm mJy$). These galaxies are referred to as "faint SMGs" hereafter. Paper I presented the number counts at flux densities  $S_{850 \mu\rm m}>0.9~\rm mJy$ and  $S_{450 \mu\rm m}>4~\rm mJy$, and an estimation of the contribution of the detected galaxies to the Cosmic Infrared Background at both wavelengths of $\sim28$ per cent. In paper II, \citet{Zavala2018} identified robust optical counterparts for 75~per cent of the galaxies and estimated their infrared luminosities, star formation rates (SFRs), dust temperatures and the median stellar mass of the population. They provided an initial comparison to the main sequence of star-forming galaxies adopting the evolution of the main sequence of a general field, and found that most faint SMGs lie within the main sequence of star formation and are predominantly disc-like galaxies, with a transition from irregular discs to discs+bulges at $z\sim1.4$, such that the bulge seems to be developing at later cosmic times. 

In this paper we further explore the properties of these faint SMGs and those of coeval optically-selected massive star-forming galaxies (hereafter referred to as SFGs) extracted from  the same field in order to address some outstanding questions that were not addressed in paper II: are faint SMGs significantly different from other massive star forming galaxies in the field? Do faint SMGs have signs of increased merging or disturbances when compared to other star-forming galaxies in the field? Does the morphological evolution detected in faint SMGs happen at the same rate as that of other massive star forming galaxies? In order to reduce biases in the comparison we consistently estimate properties for both populations extracted in the same field:  stellar mass, SFRs and morphology are characterized using the same methods for both populations. The paper is organized as follows: in section~\ref{sec:sample} we present the sample selection; in section~\ref{sec:data_source} we describe the ancillary data and catalogs we will use to explore the comparison between faint SMG and SFG populations. In section~\ref{sec:results}, we present the analysis of the star-formation main sequence, the location of faint SMGs from the main sequence and the evolution of the morphology of both populations. In section~\ref{sec:discussion} we discuss our results in the light of other results presented in the literature, and in section~\ref{sec:conclusions} we summarize the conclusions of this work.
 
We adopt the standard $\Lambda$CDM cosmology with $\Omega_{\Lambda}=0.68$, $\Omega_{m}=0.32$ and $H_{0}=67~\rm{km\, s^{-1}Mpc^{-1}}$ from \cite{Planck2014}.

\section{Sample selection}\label{sec:sample} 

\subsection{Faint Submillimeter Galaxies}
\label{sec:sampleSMG}

Our sample of faint SMGs is extracted from the 450 and $850~\mu\rm m$ catalogues of the SCUBA-2 Cosmology Legacy Survey \citep[S2CLS;][]{Geach2013,Geach2017} in the $\sim 70$ arcmin$^2$ deep survey of EGS \citep{Zavala2017}, at $\sigma_{450 \mu\rm m} = 1.2~\rm mJy/beam$ and a deeply confused instrumental $\sigma_{850 \mu \rm m} = 0.2~\rm mJy/beam$. From the initial 92 galaxies detected at $450~\mu \rm m$ and 108 galaxies detected at $850~\mu \rm m$, \citet{Zavala2018} identified robust optical or NIR counterparts for 71 of them using radio (1.4 GHz), $8~\mu\rm m$ and $24~\mu\rm m$ observations to improve on the astrometry and link the SMGs to their most probable associations. 
They estimated that 13~per cent of the faint SMGs could have incorrect counterpart associations and discarded six sources with discrepant photometric redshifts derived from optical-IR and FIR data to minimize the impact of potential incorrect identifications. They estimated the IR luminosities ($L_{\rm IR}$), FIR-based star-formation rates ($\rm{SFR_{FIR}}$) and dust temperatures of the sample using the SCUBA-2 and \textit{Herschel} photometry at the NIR or radio positions of the counterparts and analyzed the evolution of $\rm{SFR_{FIR}}$, dust temperature and morphology through time.

From these 71 galaxies with optical counterparts, we further restrict the main sample of study to 57 galaxies which fall within the footprints of the deep {\it Hubble Space Telescope} ({\it HST}) imaging where  our comparison sample of optically selected star-forming galaxies is extracted. The dusty galaxies have a median $S_{850 \rm\mu m} = 2.0 \pm 1.2~\rm mJy/beam$ and $L_{\rm IR} = 10^{12.0\pm 0.5}~\rm L_{\odot}$, such that 37~per cent have luminous infrared galaxy (LIRG) luminosities ($10^{11}~\rm{ L_{\odot} }$ $\leq L_{\rm IR} < 10^{12}~\rm{L_{\odot}}$), 54~per cent have ultraluminous infrared galaxy (ULIRG) luminosities  ($10^{12}~\rm{L_{\odot}}$ $\leq L_{\rm IR} < 10^{13}~ \rm{L_{\odot}}$) and 9~per cent have luminosities below the LIRG regime ($L_{\rm IR} < 10^{11}~\rm{L_{\odot}}$).

\subsection{Comparison sample of star-forming galaxies} 
\label{sec:sample_select}

We extract a comparison sample of optically selected SFGs from the EGS field mapped by the Cosmic Assembly Near-infrared Deep Extragalactic Legacy Survey \citep[CANDELS -][]{Grogin2011,Koekemoer2011}. 
\citet{Stefanon2017} presented the catalogue of 41,457 sources detected in the \textit{H}-band (F160W) within the deep $\sim206~\rm{arcmin}^{2}$ footprint of the survey. The 3D-HST catalogue \citep{Skelton2014,Momcheva2016} of the field contains 41,200 objects. We cross-matched the two catalogues using a search radius of $r= 0.5~\rm{arcsec}$. 

\begin{figure*}
    \includegraphics[width=\textwidth]{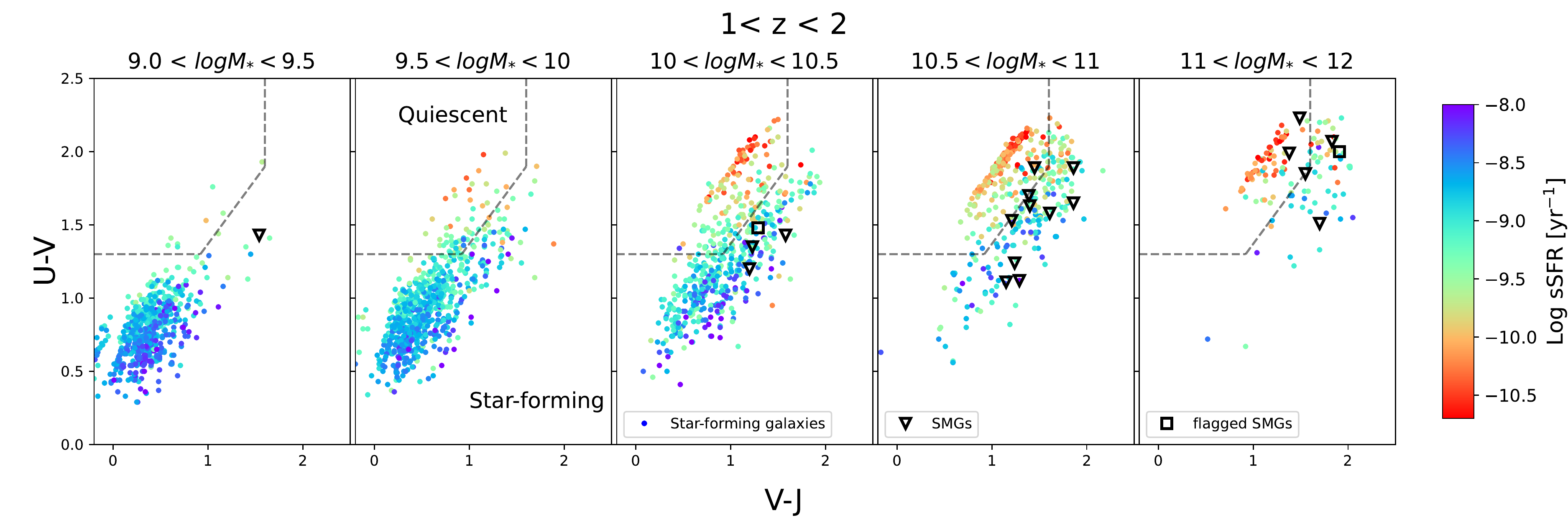}
    \caption{\textit{UVJ} color-color diagram of $H$-band selected galaxies at $1<z<2$, where we use rest-frame AB magnitudes, not corrected for dust extinction. The panels separate the galaxies in narrow bins of stellar mass ($\Delta \log (M_{\star}/\rm{M_{\odot}}))=0.5$). The galaxies are color-coded according to their sSFR (sSFR=SFR/$M_{\star}$), adopting the SFR that \citet{Barro2019} estimated with FIR, NIR or UV data and the stellar masses from \citet{Stefanon2017}. The faint SMGs with robust optical counterparts are represented with empty symbols, indicating whether they comply with the good quality selection criteria applied to SFGs (triangles), or instead they were flagged as not complying (squares). The dashed lines divide the quiescent (upper left) and star-forming regions \citep{Williams2009}. All galaxies outside of the quiescent region that are not identified as SMG counterparts are included in our optically-selected star-forming comparison sample. Eighteen per cent of faint SMGs lie within the quiescent region. These are hence heavily dust-obscured systems that the \textit{UVJ} diagram is not able to properly classify as star-forming galaxies.}  
    \label{UVJ_M_z}
\end{figure*}

The SFG sample is defined following the method outlined by \citet{Fang2018}. We apply the following selection criteria: (i) magnitude in \textit{H}-band < 24.5 mag to ensure good \texttt{GALFIT} fittings; (ii) SExtractor parameter \texttt{CLASS\_STAR} <0.9 to avoid contamination by stars; (iii) Photometric (PhotFlag=0) and structural parameter (\texttt{GALFIT} flag=0) quality flags to exclude spurious sources and ill-constrained \texttt{GALFIT} estimations. We then use the rest-frame $U-V$ vs. $V-J$ diagram to discriminate between star-forming and quiescent galaxies.  Figure~\ref{UVJ_M_z} shows the color-color diagram for galaxies at $1.0<z<2.0$ and Appendix~\ref{UVJ_ext} contains the diagrams for all redshiff bins. We segregate star-forming and quiescent galaxies using the loci defined by \citet{Williams2009}. At $z>2.0~$ they found that galaxies showed a relation between the rest-frame $V-J$ color and $24~\mu\rm m$ flux densities, which implies that their selection criteria, despite being incomplete and classifying some red star-forming systems as quiescent, is still extendable to higher redshifts. Therefore, for $z>2.5$ we use the division relationships as for $2.0<z<2.5$. 

The SFG sample contains all $UVJ$-classified star-forming galaxies that are not counterparts of SMGs:  4878 sources at $0.2<z<4$. This sample will allow us to derive the properties of faint SMG counterparts and coeval optically-selected star forming galaxies in the same way, to assess if faint SMGs are unique in any way among star-forming galaxies.

\section{Ancillary data}
\label{sec:data_source}

\subsection{Stellar Masses}

The CANDELS catalogues\footnote{ \url{ https://archive.stsci.edu/missions/hlsp/candels/egs/catalogs/v1/} } present stellar masses estimated by 10 independent teams. The methodologies are described by \citet{Santini15}, who found that all methods that used the same stellar population templates are in overall good agreement. \citet{Mobasher} also showed a comparison of the stellar masses independently derived by the different teams, finding no significant bias between them and a similar scatter of $\sigma(\log M_{\star}/\rm{M_{\odot}})=0.136~\rm dex$. 

We adopt the stellar masses of the 2a\_tau team, who also presents star formation rates. The stellar masses were estimated fitting spectral energy distributions (SED) with the \texttt{FAST} code \citep{Kriek2009}, a \cite{Chabrier} Initial Mass Function (IMF), \cite{BruzualCharlot} stellar population templates and the \cite{Calzetti} extinction law.

\subsection{Redshifts}
\label{sec:z_source}

We will use 3D-HST redshifts \citep{Momcheva2016} for the SMG and SFG sample. The mean difference between 3D-HST and CANDELS photometric redshifts is $0.013\pm0.003$. All SMGs have optical-IR photometric redshifts consistent with FIR photometric redshifts \citep{Zavala2018}, as this was used as a criterium to exclude counterparts that were possible misidentifications (see section~\ref{sec:sampleSMG}).

\subsection{Star Formation Rates}
\label{sec:SFR_source}

The SFRs estimated by the 2a\_tau team use a combination of ultraviolet (UV) and IR-based tracers \citep{Barro2019}: 
\begin{itemize}
    \item $\rm{SFR_{UV}}$, the SFR derived from the 2800~\r{A} rest-frame flux density, not accounting for dust extinction. 
    \item $\rm{SFR_{UV}^{corr}}$, the SFR derived from the 2800~\r{A} rest-frame flux density, corrected by dust extinction.
    \item $\rm{SFR_{UV+IR}^{W11}}$, the SFR extrapolating the  \textit{Spitzer}/MIPS $24~\mu\rm m$ emission with the SFR template by \citet{Wuyts2011_SFRs}, and co-adding SFR$_{\rm UV}$, $\rm{SFR_{UV+IR}^{W11}}= \rm{SFR_{UV}} + \rm{SFR^{W11}_{IR}}$.
    \item  $\rm{SFR_{UV+IR}^{Herschel}}$, the SFR derived from fitting dust emission templates to
    \textit{Herschel} photometry,  and co-adding the SFR$_{\rm UV}$,  $\rm{SFR_{UV+IR}^{Herschel}= {SFR_{UV}} + {SFR}^{Herschel}_{\rm IR}}$.
\end{itemize}

These estimates are available for both the SFG and SMG sample.

We also use in our analysis the $\rm{SFR}_{FIR}$ estimated by  \citet{Zavala2018}, where they fitted a modified black body with fixed emissivity index $\beta=1.6$ at the redshift of the 3D-HST catalogues \citep{Skelton2014,Momcheva2016} to estimate $L_{\rm IR}$ and then calculated SFR$_{\rm FIR}$ using the \cite{Kennicutt1998} calibration with a \citet{Chabrier} IMF. 

Since source 850.26 has no $L_{\rm IR}$ reported in \cite{Zavala2018}, but has a robust optical counterpart, in this paper we estimate $L_{\rm IR}$ for this galaxy in a similar manner. We fitted a modified black body with the median dust temperature of the sample $T_{\rm{d}}=45~\rm K$ and a fixed emissivity index $\beta=1.6$ at the 3D-HST redshift \textbf{($z=2.52 \pm 0.01$)}, obtaining $L_{\rm{IR}} = 10^{12.16\pm0.08}~\rm L_\odot$ and $\rm{SFR_{FIR}}=150\pm30~\rm M_{\odot}~yr^{-1}$, which is within the values found for the rest of the SMG sample.

\subsection{Dust extinction}

The dust extinction $A_{V}$ presented by \citet{Barro2019} is estimated from the slope of the UV continuum, $\beta_{\rm UV}$, and then corrected using the IRX-$\beta_{\rm UV}$ relation to account for multiple attenuation laws, where IRX is the infrared excess derived from the ratio between  IR-based and UV-based SFRs. They found higher IRX values for galaxies with higher $\rm{SFR_{IR}}$ at the same $\beta_{\rm UV}$ value, where galaxies with $\rm SFR_{IR}>70~\rm M_{\odot}yr^{-1}$ lie above the IRX-$\beta_{\rm UV}$ relation.

\subsection{\textit{H}-band morphologies}

We adopt the structural parameters derived by \citet{vdWel2014} from the {\it HST} $H$-band images.
Using an automated process \citet{VdW2012} fitted a S\'ersic model with \texttt{GALFIT}  \citep{Galfit} to the galaxies in the field.

For those SMG counterparts with flagged \texttt{GALFIT} fits or outlier structural parameters, we replaced the fits of \citet{vdWel2014} by those of \citet{TesisKarla}, who also performed \texttt{GALFIT} fits, individually, carefully masking out any companion galaxies and nearby stars. 

We also use the visual morphological classification presented by \citet{HuertasCompany}, where they trained a Convolutional Neural Network (CNN) with the visual classification of galaxies in the GOODS-S field by \citet{Kartaltepe15}. \citet{HuertasCompany} then classified the galaxies in all the CANDELS fields using the same method.

\section{Analysis: comparison between submillimeter and optically selected star forming galaxies}
\label{sec:results}
\subsection{Stellar masses}
\label{sec:analysis_stellarmass}

\begin{figure*}
    \includegraphics[width=\textwidth]{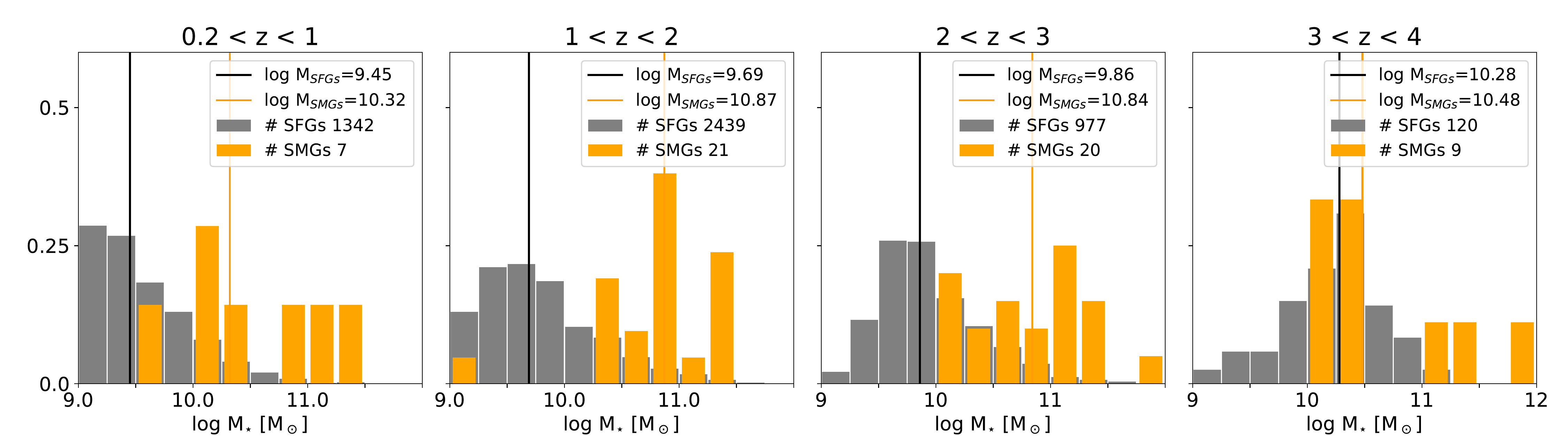}
    \caption{ Normalised distributions of stellar mass for optically-selected star-forming galaxies (SFGs, grey) and optical counterparts of submillimeter galaxies (SMGs, orange). The median stellar masses (vertical lines) of faint SMGs are  significantly higher than those of SFGs for all redshift bins.}
    \label{fig:hist_M}
\end{figure*}

\begin{table}
    \begin{center}
    \caption{Median stellar mass of the optically-selected star-forming galaxy (SFG) and faint submillimeter galaxy (SMG) samples. The columns are (1) redshift range; (2) median stellar mass of SFGs; (3) median stellar mass of faint SMGs. The errors of the median values of stellar mass were calculated with a bootstrap.}
    \begin{tabular}{ccccc}
        \hline
        $z$ & $\log M_{\star,\rm{SFGs}}$ & $\log M_{\star,\rm{SMGs}}$ \\ 
        & [$\rm M_{\odot}$] & [$\rm M_{\odot}$] \\ \hline
        0.2--1.0 & $9.46\pm0.01$ & $10.3\pm0.6$ \\ 
        1.0--2.0 & $9.70\pm0.01$ & $10.9\pm0.1$ \\ 
        2.0--3.0 & $9.86\pm0.02$ & $10.8\pm0.2$ \\
        3.0--4.0 & $10.28\pm0.05$ & $10.5\pm0.1$ \\ \hline 
        all $z$ & $9.68\pm 0.01$ & $10.8\pm0.1$ \\ \hline 
    \end{tabular}
    \end{center}
    \label{tab:med_mass}
\end{table}

The mean stellar mass of the faint SMG sample studied in this work is $\log (M_{\star}/\rm{M_{\odot}})=10.75\pm0.07$. This is slightly lower than the $\log (M_{\star}/\rm{M_{\odot}})=10.95\pm0.03$ value estimated by \citet{Zavala2018}, who adopted the \citet{Momcheva2016} 3D-HST catalogue values. The difference between these estimates is mainly driven by 4 galaxies without estimations of stellar masses in the  3D-HST catalogue, and 3 galaxies with different redshift estimations between the CANDELS and 3D-HST catalogues, which result in mass differences  $\Delta (\log (M_{\star}/\rm{M_\odot}))=1-2~\rm dex$.

The distributions of the stellar masses of SFGs and SMGs in 4 redshift bins are presented in Figure~\ref{fig:hist_M}, and their medians are listed in Table~\ref{tab:med_mass}. Errors in the medians are estimated through a bootstrap analysis. The median masses of SMGs are systematically larger than those of SFGs. In order to assess the statistical significance of this claim, we apply a Mann-Whitney test \citep{Mann-Whitney} to analyse whether the medians of both populations could be compatible with that of a single parent distribution. For all redshift bins the probabilities for the null hypothesis to be true are $p<0.05$ (see Table~\ref{tab:med_massApp}), and hence we reject the null hypothesis of statistical identity of the medians. We use the uncertainty in the redshift estimations in the 3D-HST catalogue to assign to each galaxy a new redshift and recalculate the test 5000 times, finding that the result is robust ($p \sim 5\times 10^{-3}$ -- $3\times 10^{-13}$ at different redshift bins). We hence conclude that faint SMGs typically have higher masses than optically selected star-forming galaxies, as has been discussed in the literature for classical SMGs  \citep[e.g. ][]{Blain2004, Chapman2005, MinYun2012}. 

We note that we do not have sufficient statistical significance to claim an increment in the stellar mass of SMGs from the redshift bin $0.2<z<1$ to $1<z<3$. In contrast, the SFG population shows a consistent increase in stellar mass with redshift, due to Malmquist bias. 

In order to check if dust extinction could bias our results, we select a sub-sample of SFGs with similar dust extinction $A_{V}$ and $V-J$ colors as the SMG sample: $A_{V}=1.4\pm0.09$, $V-J=1.38^{+0.03}_{-0.05}$. We still find the statistical difference between stellar masses of the SMG and SFG samples at  $0.2<z<3$. At $3<z<4$, however, the null hypothesis of identity cannot fomally be rejected ($p=0.25$). Hence, color selection and dust extinction, as measured from optical-infrared data, cannot account for the differences in mass found between the SFG and SMG samples.

\subsection{Star Formation Rates}
\label{sec:SFR}

\subsubsection{Star-formation main sequence}
\label{sec:SFMS}

\begin{figure*}
    \includegraphics[width=\textwidth]{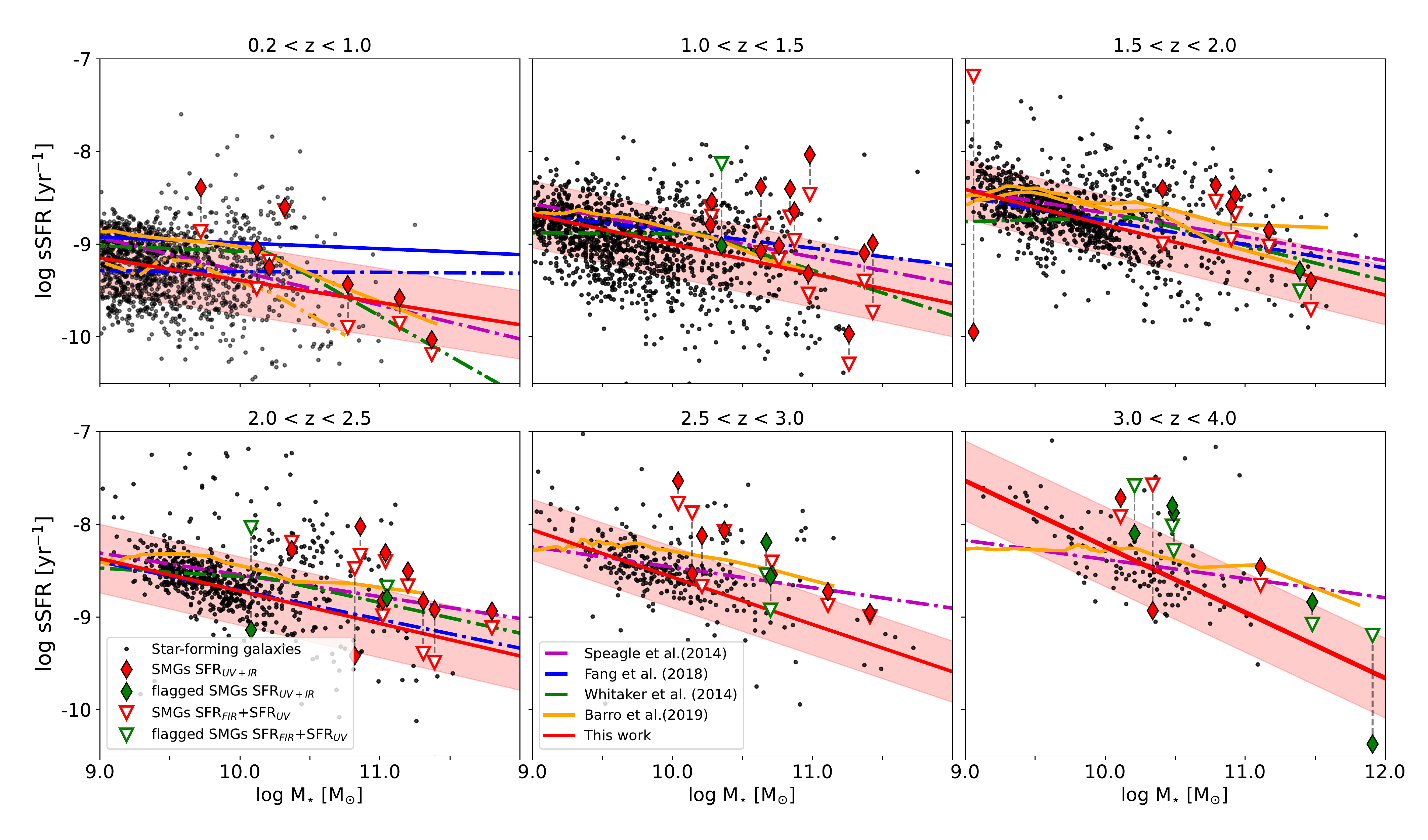}
    \caption{Specific SFR vs. $M_{\star}$ for the SFG sample (black dots), adopting the best-estimate of SFR ($\rm{SFR_{UV}^{corr}}, \rm{SFR_{UV+IR}^{W11}}$ or $\rm{SFR_{UV+IR}^{Herschel}}$) by \citet{Barro2019} for SFGs. Red filled diamonds represent the sSFRs based on $\rm{SFR_{IR} +SFR_{UV}}$ estimations for the SMGs with optical-NIR counterparts that comply with the good quality selection criteria in section~\ref{sec:sample_select}. The empty red triangles connected with them by dashed lines show the corresponding sSFRs based on $\rm{SFR_{FIR+UV}}$. Similarly, green filled diamonds and empty triangles represent the estimates for SMGs that were flagged out by the optical-NIR quality selection criteria. The red solid line is the star-formation main sequence fit for SFGs and the shaded region the 1$\sigma$ scatter. We also show the main sequences derived by \citet[][blue]{Fang2018}, \citet[][purple]{Speagle2014}, \citet[][green]{Whitaker2014} and \citet[][orange]{Barro2019}. At redshift bins $0.2<z<1.0$ and $1.5<z<2.0$ we represent the main sequences derived by \citep{Fang2018} and \citep{Barro2019} in their own bin definitions: $0.0 < z< 0.5$, $0.5 < z < 1.0$, $1.4 < z < 1.8$ and $1.8 <z< 2.2$ for \citet{Barro2019}, and $0.2<z<0.5$ and $0.5<z<1.0$ for \citep{Fang2018}. We find that at $z<2.5$ the main sequences derived in the literature agree well with ours within the $1\sigma$ scatter, and at $2.5<z<4$ our slope is steeper. }
    \label{fig:SFMS}
\end{figure*}

The tight correlation between SFR and $M_{\star}$ for SFGs is referred to as the main sequence of star-forming galaxies \citep{Noeske2007, Daddi2007, Elbaz2011, Whitaker2014, Speagle2014}. This is often expressed in terms of the specific star formation rate, sSFR=SFR/$M_{\star}$. Figure~\ref{fig:SFMS} shows the main sequence estimated for our optically-selected SFG sample in 6 redshift bins using the best SFR in the CANDELS catalog available for each source: $\rm{SFR_{UV+IR}^{Herschel}}$,  $\rm{SFR_{UV+IR}^{W11}}$, or $\rm{SFR_{UV}^{corr}}$, in that order of preference, which are collectivelly denoted as $\rm{SFR_{IR+UV}}$. We fitted the $\log{\rm sSFR}-\log M_{\star}$ relation with a linear function and an iterative 3-step least-squares method, using two $1.5\sigma$ clippings on the surviving sample. The parameters of the main sequence fits are listed in Table~\ref{tab:SFMS_fit}. 

We note that our main sequence is mostly consistent, within the RMS, with previously derived main sequences at $z<2.5$ ~\citep[i.e.][]{Speagle2014, Whitaker2014, Fang2018, Barro2019}, despite the fact that these comparison main sequences were derived using samples extracted at different depths, with different SFR estimations and functional forms for the fit (e.g. power law, broken power law, mass-time dependant, etc). At redshifts $z>2.5$, however, our main sequence fit is steeper than the main sequences derived by \citet{Speagle2014} and \citet{Barro2019}. 

\begin{table}
    \begin{center}
    \caption{Parameters of the linear fits $\log ({\rm sSFR/yr^{-1}}) =m~\log (M_{\star}/ {\rm M_{\sun}}) + b$. The columns give: (1) redshift bin, (2) slope, (3) zero point, (4) RMS of the SFR of the galaxies to the best fit sequence at their corresponding stellar mass.}
    \begin{tabular}{c|ccc}
        \hline  
        $z$ & $m$ & $b$ & $\sigma$ \\  \hline
        0.2--1.0 & $-0.24$ & $-6.99$ & 0.37 \\
        1.0--1.5 & $-0.32$ & $-5.80$ & 0.36 \\
        1.5--2.0 & $-0.38$ & $-4.99$ & 0.32 \\
        2.0--2.5 & $-0.35$ & $-5.22$ & 0.37 \\
        2.5--3.0 & $-0.51$ & $-3.47$ & 0.33 \\
        3.0--4.0 & $-0.71$ & $-1.14$ & 0.43 \\ \hline
    \end{tabular}
    \end{center}
    \label{tab:SFMS_fit}
\end{table}

\subsubsection{Location of SMGs with respect to the Main Sequence}
\label{sec:SMGlocMS}

\begin{figure*}
    \includegraphics[width=0.8\textwidth]{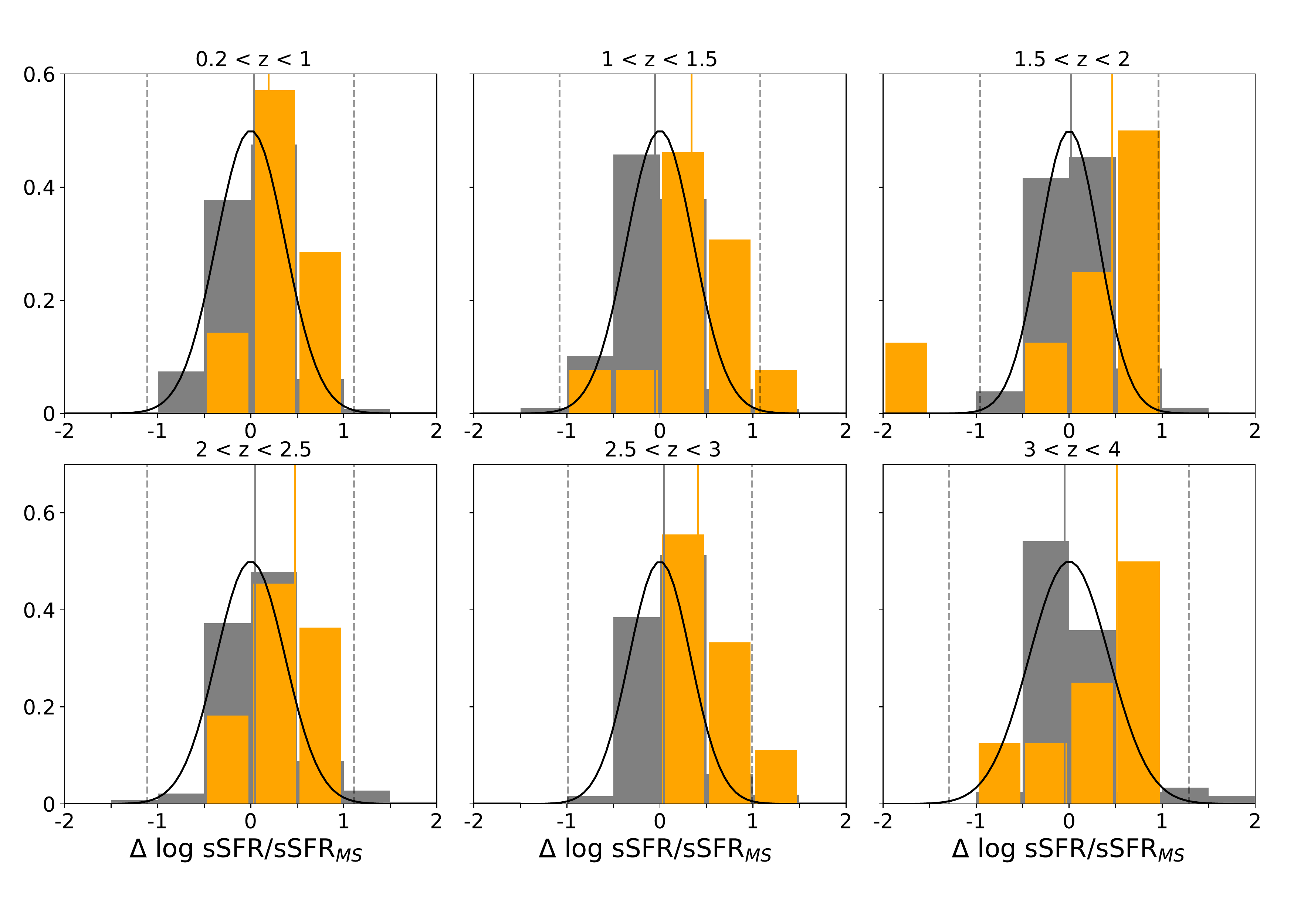}
    \caption{Distribution of the sSFR with respect to the star-formation main sequence for SMGs (orange) and SFGs (grey). We use $\rm{SFR_{IR}+SFR_{UV}}$ for the SMGs and SFGs. The Gaussian distribution marked by the solid line is centered on the main sequence for each redshift bin and the vertical dashed lines mark the $3\sigma$ limits. The median $\Delta \log \rm{sSFR}=\log (\rm{sSFR/sSFR_{MS}})$ of SMGs (orange vertical line) is significantly larger at $1<z<4$. The redshift bin with the highest fraction of SMGs above $1\sigma$ is $2.5<z<3$.}
    \label{fig:hist_SFMS}
\end{figure*}

Adopting the $\rm{SFR_{IR+UV}}$ estimations for SMGs of the CANDELS catalogs we find that 82 per cent of SMGs (42 galaxies) are located above the main sequence of SFGs, 56 per cent of SMGs (32 galaxies) are located above $1\sigma$, 21~per cent of SMGs (12 galaxies) are above $2\sigma$, and 4~per cent (2 galaxies) are located above $3\sigma$ (see Figure~\ref{fig:SFMS}). On the other hand, 11 per cent of SMGs (6 SMGs galaxies) are below $-1\sigma$ of the main sequence and one galaxy below $-2\sigma$ of the main sequence. 

In Figure~\ref{fig:SFMS} we can observe that at all redshifts faint SMGs are located at higher sSFRs than SFGs for the same stellar mass, indicating more vigorous star forming activity across the faint SMG population. Figure~\ref{fig:hist_SFMS} shows the normalized distribution of the sSFR differences to the main sequence for SMGs and SFGs, $\Delta \log \rm{sSFR}=\log (\rm{sSFR/sSFR_{MS}})$ to highlight this effect. We find positive median values of $\Delta \log \rm{sSFR}$ for the SMG sample at all redshifts. We applied the Mann-Whitney test to check if the medians of SMGs could be derived from the same parent distribution as those of SFGs, and we find the differences to be robust at $1<z<4$, once we consider the uncertainties on redshift estimations through a bootstrap: $p\sim 0.009 - 7 \times 10^{-10}$ at different redshift bins (see Table~\ref{tab:med_deltasfr_iruv} in the Appendix).

We observe at $2.5<z<3$ the highest fraction of SMGs above the main sequence: $\sim$89~per cent of SMGs (8/9) have sSFRs above $1\sigma$ of the main sequence. This is also the redshift bin where the $850~\mu\rm m$-selected SMG population and the dust-obscured SFR density peak \citep[$z\sim2-2.5$, e.g.][] {Zavala2021}. At higher redshift bins, $3<z<4$, the faint SMGs do not have larger SFRs than the SFGs and we see a larger dispersion of faint SMGs across the main sequence.

\citet{Elbaz2011} initially used $R_{\rm SB}=\rm sSFR/sSFR_{MS}\geq2$ as the threshold for defining a starburst. Nowadays a factor of 3 is more commonly used for this \textit{starburstiness} parameter \citep{Franco20}. If we consider this threshold, we would classify 23 (40~per cent) of the faint SMGs as starburst galaxies.

We applied the same colour and dust extinction selection as in section~\ref{sec:analysis_stellarmass} to define a sub-sample of SFGs with the same dust extinction properties as SMGs, to check if this could bias our results. The fraction of starbursts and location of SMGs with respect to the main sequence of SFGs remains unchanged.

We hence conclude that the faint SMG population is mostly located above the main sequence of star formation (56~per cent above $1\sigma$), particularly at $z>2$ where the population peaks (65~per cent). There is also a significant fraction of starbursts (40~per cent).

\subsubsection{Impact of different SFR estimates}
\label{sec:SFRcompa}

\begin{figure}
    \includegraphics[width=0.45\textwidth]{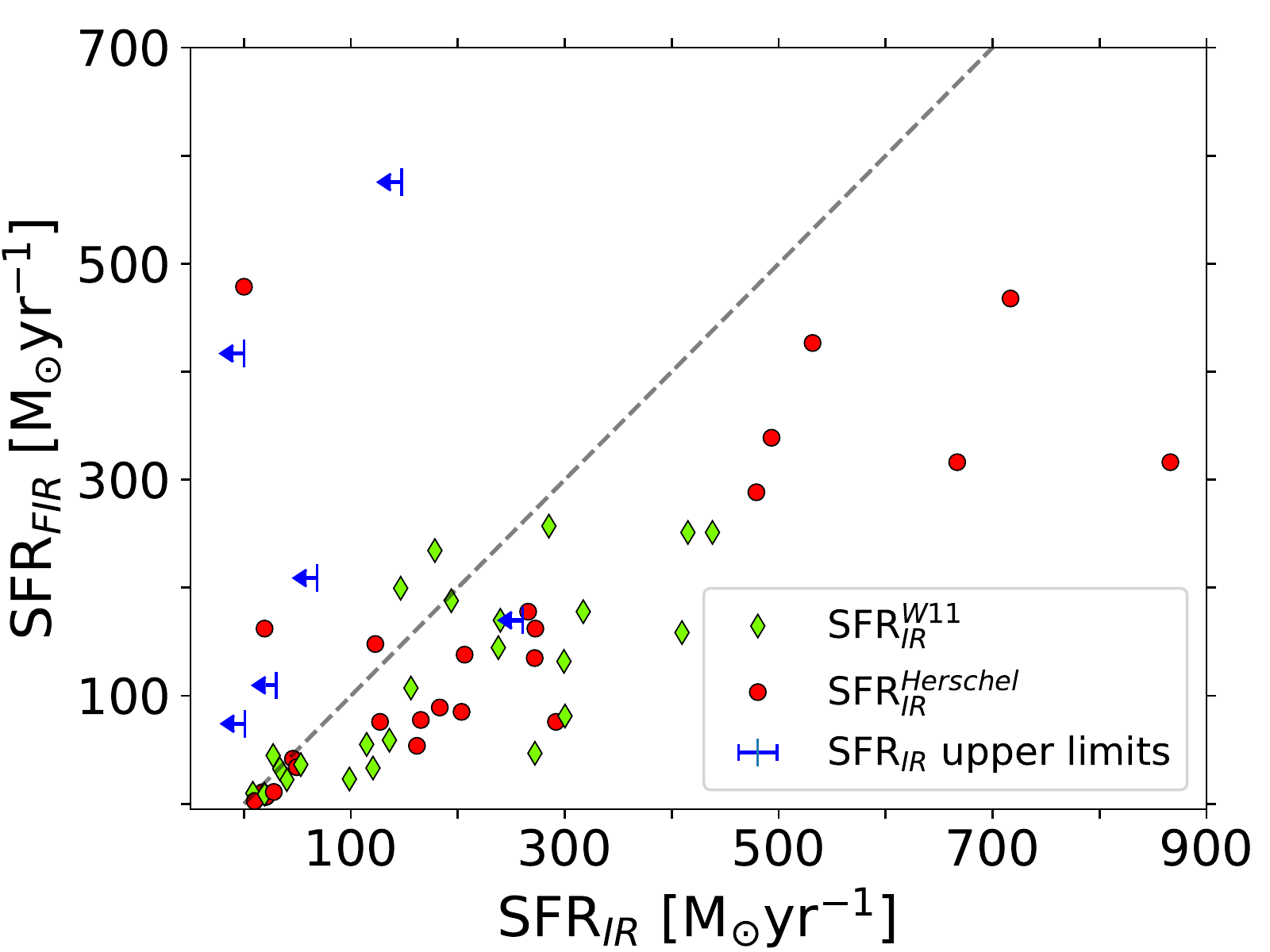}\\
    \includegraphics[width=0.5\textwidth]{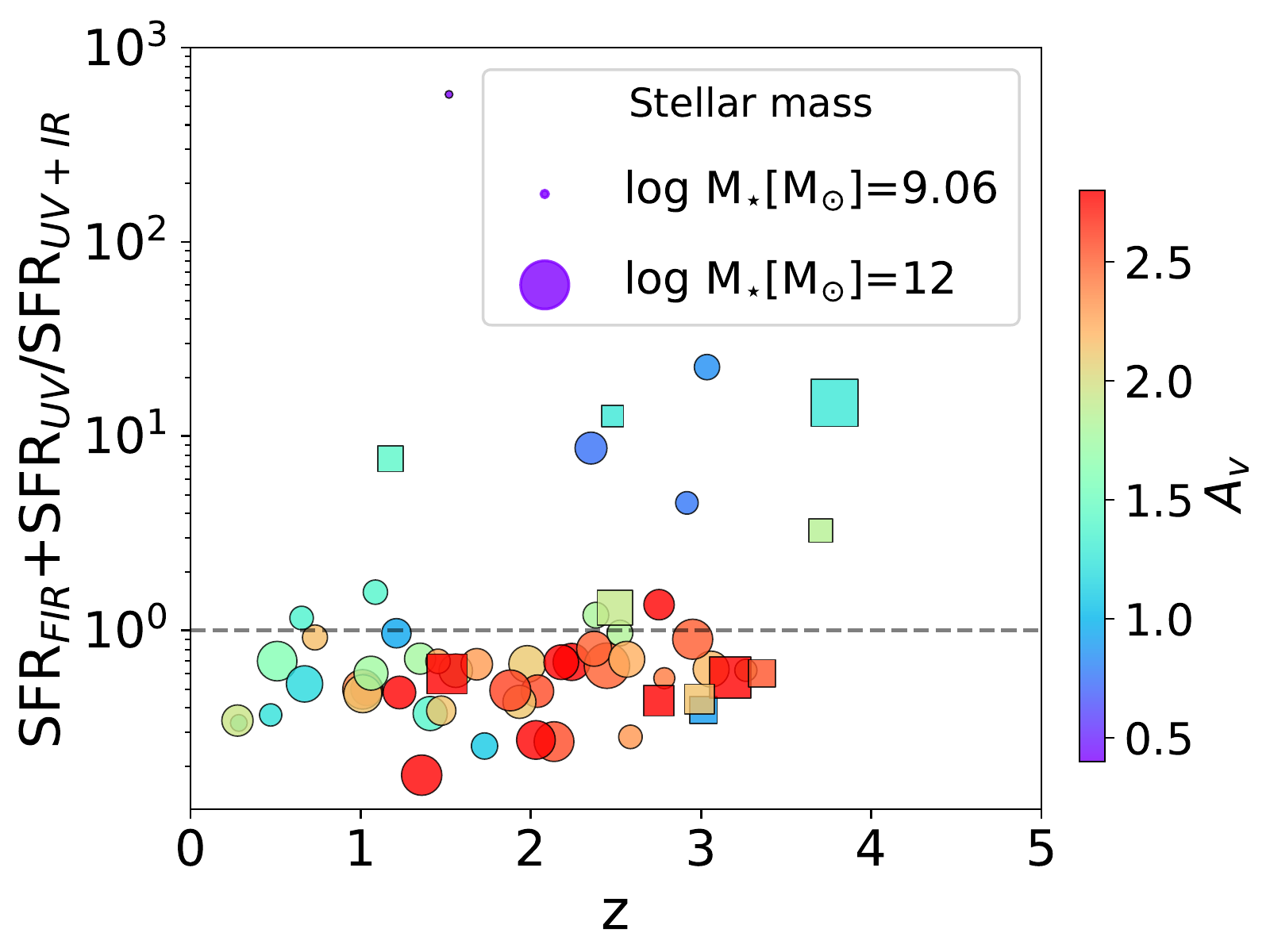}
    \caption{
    (Top panel) Comparison of IR-derived SFRs: $\rm{SFR_{FIR}}$ by the S2CLS team \citep{Zavala2018}, using the 450/850~$\mu$m SCUBA-2 and \textit{Herschel} deconvolved photometry at the NIR or radio positions of the counterparts and $\rm{SFR_{IR}}$ by the CANDELS team \citep{Barro2019}, using either a fit to the \textit{Herschel} photometry extracted with a PSF or, whenever that was not available, extrapolating the \textit{Spitzer}/MIPS 24~$\mu$m photometry with templates that extend to the FIR. The arrows indicate the upper limits derived from \textit{Herschel} data for the CANDELS catalog whenever a detection at 24~$\mu$m or at longer wavelengths was not available.
    (Bottom panel) SFR ratio versus redshift, dust extinction (color gradient) and stellar mass (sizes). The color in the symbols indicate the dust extinction $A_V$ and the sizes are proportional to the stellar mass $M_{\star}$ of each galaxy. We use the total SFRs coadding the IR SFR and raw UV-based SFR considering both FIR and IR-derived SFR estimations. When the SFR is estimated without any IR data it uses the UV SFR corrected for dust extinction, which in the cases of great discrepancy shows small values. } 
    \label{fig:SFR_IR_FIR} 
\end{figure}

We explore how different estimations of the total SFRs for faint SMGs impact their location with respect to the main sequence. The top panel of Figure~\ref{fig:SFR_IR_FIR} compares the SFRs based on IR observations derived by the CANDELS \citep[$ \rm{SFR_{IR}},$ ][]{Barro2019} and the S2CLS teams \citep[$\rm{SFR_{FIR}} $,][]{Zavala2018}.

There are six SMGs that did not have \textit{Herschel} nor \textit{Spitzer} detections at the time the CANDELS catalog was produced (marked with blue arrows in Figure~\ref{fig:SFR_IR_FIR}). These galaxies have higher values of $\rm{SFR_{FIR}}$ than the upper limits of $\rm{SFR_{IR}}$ derived from \textit{Herschel} non-detections. In these cases, the best SFR estimate uses the UV SFR corrected for dust extinction, but this dust correction fails to recover the total SFR from the UV continuum solely. 
These are also galaxies with $A_V$ values unusually small (tipically $A_V<1$), compared to the  obscuration values $A_V>2$ of the bulk of the faint SMG optical counterparts. This can be appreciated in the bottom panel of Figure~\ref{fig:SFR_IR_FIR}, where the SMGs with larger discrepancies between the SFR estimations also have low dust extinction values. This effect of small derived obscuration but high $\rm SFR_{FIR}$ could be explained in a patchy dust distribution scenario, where most of the UV light would come from areas of less obscuration, hence rendering an underestimation of dust obscuration for the full galaxy. The S2CLS $\rm{SFR_{FIR}}$ is in these cases higher than the upper values derived from {\it Herschel} photometry, as they are based on the detections at 450/850~$\mu$m at a higher spatial resolution and the deconvolved {\it Herschel} fluxes at the position of the IR and radio counterparts, providing a better characterization of the dust-enshrouded star formation rate.

For most other galaxies $\rm{SFR_{FIR}}$ are similar or smaller than $\rm{SFR_{IR}}$. We note, however, that $\rm{SFR_{FIR}}$ was estimated by deconvolving the \textit{Herschel} flux densities at the position of the radio or mid-IR counterpart sources \citep{Zavala2018}. Meanwhile,  $\rm{SFR_{IR}}$ is based on \textit{Herschel} flux densities extracted with a PSF model, without deconvolution \citep{Barro2019}. Hence $L\rm{_{IR}}$ is likely boosted by the crowding and merging of fainter sources into the main SMG extracted flux density, and $\rm{SFR_{IR}}$ could be overestimated. 

\begin{table}
    \centering
    \caption{Median SFRs of the faint SMGs in our sample, considering various sets of data and estimation methods. The medians of $\rm{SFR_{FIR}}$ and $\rm{SFR_{IR}}$ agree with each other within the errors. } 
    \begin{tabular}{|c|c|}
        \hline
         $\rm{SFR_{FIR}}$ & $132\pm28~\rm{M_{\odot}~yr^{-1}}$ \\
         $\rm{SFR_{IR}}$ & $151\pm33~\rm{M_{\odot}~yr^{-1}}$ \\
         $\rm{SFR_{UV}}$ & $2.8\pm0.5~\rm{M_{\odot}~yr^{-1}}$ \\\hline
         $\rm{SFR_{FIR} +SFR_{UV}}$ & $133\pm27~\rm{M_{\odot}~yr^{-1}}$ \\
         $\rm{SFR_{UV+IR}}$ & $152\pm30~\rm{M_{\odot}~yr^{-1}}$ \\ 
         \hline
    \end{tabular}
    \label{tab:med_sfrs}
\end{table}

In Table~\ref{tab:med_sfrs} we present the median SFRs for the faint SMG sample based on the estimates presented in section~\ref{sec:SFR_source}. We also present a total SFR for faint SMGs by coadding the $\rm{SFR_{FIR}}$ calculated by \citet{Zavala2018} and the raw UV-based SFR not corrected for dust obscuration of \citet{Barro2019}:  $\rm{SFR_{FIR} +SFR_{UV}}$. This estimate is also represented in Figure~\ref{fig:SFMS}. The median values of $\rm{SFR_{FIR} +SFR_{UV}}$ and $\rm{SFR_{UV+IR}}$ are in agreement with each other within the errors. The ratio of median SFRs is $\langle\rm{SFR_{FIR} +SFR_{UV}}\rangle/ \rm{\langle SFR_{UV+IR}}\rangle = 0.9\pm0.2 ~\rm M_{\odot}~yr^{-1}$ and the median of the ratios $\langle \rm{SFR_{FIR} +SFR_{UV}}/ \rm{SFR_{UV+IR}} \rangle = 0.59\pm0.04$, highlighting the overall tendency for $\rm{SFR_{FIR} +SFR_{UV}} < \rm{SFR_{UV+IR}}$ in our sample. The SMGs that do not follow this tendency are mainly those with $\rm SFR_{UV+IR}$ estimations derived from UV-based SFRs corrected by dust extinction. These discrepant SMGs also show low values of $A_V$ (see Fig.~\ref{fig:SFR_IR_FIR}).

Adopting the $\rm{SFR_{FIR}+SFR_{UV}}$ estimations for SMGs, the location of faint SMGs in the sSFR vs $M_{\star}$ diagram is such that 43 galaxies (75 per cent) are above the main sequence of SFGs, 29 SMGs (51~per cent) are located above $1\sigma$, 7 SMGs (12~per cent) are above $2\sigma$, and only 1 (2~per cent) is located above $3\sigma$. On the other hand, there are  4 SMGs (7~per cent) below $-1\sigma$ and one galaxy below $-2\sigma$ of the main sequence. Hence the number of starbursts is slightly reduced from those adopting the CANDELS SFR estimates, but they still are indicative of  higher SFRs than the optically selected population of SFGs of the same stellar mass. We confirm this statement with a Mann-Whitney test, finding that at $1<z<4$ the sSFR of SMGs is larger than that of SFGs ($p=10^{-3} - 10^{-9}$).

\subsection{\textit{H}-band Morphology}
\label{sec:struct}

We explore the morphological differences between faint SMGs and SFGs in \textit{H}-band  using the structural parameters derived by \citet{vdWel2014}. We selected the SFG sample to have only good \texttt{GALFIT} fits with flag=0. Since \citet{vdWel2014} estimated that $\sim 5$~per cent of the sample had catastrophic or bad fits, errors in size, redshift or stellar mass, and our SMG sample is small, we reviewed the \textit{H}-band postage stamps of all SMG counterparts and individually evaluated whether we could accept the morphological parameters provided in the catalogue. We specially examined the cases with bad \texttt{GALFIT} flags, cases with very small or large sizes ($R_{\rm e}<0.6~\rm kpc$ or $R_{\rm e}>10~\rm kpc$) or S\'ersic indices close to the constraints introduced in the automated process ($n=0.2$ and $n=8$). We replaced the structural parameters of the outlier galaxies with those estimated by \citet{TesisKarla}, who also fitted them with a single S\'ersic profile using \texttt{GALFIT}, after masking out any companion galaxies or nearby stars, and found reduced-$\chi^2$ values similar to those in \citet{vdWel2014}.
The following galaxies have revised structural parameters: 
\begin{itemize}
  \item 850.028 at $z=2.5^{+0.4}_{-0.1}$ lies close to the diffraction spike of a field star. The catalogue shows a large radius of $R_{\rm e}=23\pm15~\rm kpc$ and disk-like morphology with index $n=1.06\pm1.22$, and a good-fit flag=0. We adopt the more moderate radius $R_{\rm e}=4.8~\rm kpc$ and $n=1.23$. 
  \item 850.007 lies at $z=3.0\pm0.1$ and the catalogue includes a very large radius $R_{\rm e}=31\pm17~\rm kpc$ and S\'ersic index $n=8\pm 6$ with a bad fitting flag. We adopt $R_{\rm e}=1.97~\rm kpc$ and $n=0.68$. 
  \item 850.030 and 850.069 are located at $z=1.52\pm 0.06$ and $z=0.560\pm0.001$, and have very close companions and perturbed morphologies. Their morphological parameters in the catalogue are $R_{\rm e}=21.5\pm0.3~\rm kpc$,  $n=1.10\pm0.03$ and $R_{\rm e}=11.1\pm0.3~\rm kpc$,  $n=1.310\pm 0.005$, respectively. We use instead the more conservative values $R_{\rm e}=14.8~\rm kpc$, $n=1.25$ and $R_{\rm e}=9.2~\rm kpc$, $n=0.71$, respectively. 
  \item 850.044 ($z=3.17\pm0.14$), 850.56 ($z=2.48\pm0.01$) and 850.072 ($z=3.36\pm0.35$) have bad fitting flags in the catalogue. We adopt $R_{\rm e}=2.48~\rm kpc$, $n=0.45$; $R_{\rm e}=8.46~\rm kpc$, $n=0.49$; and $R_{\rm e}=1.83~\rm kpc$, $n=0.32$, for these cases, respectively.
\end{itemize}

Table~\ref{tab:med_params} lists the median values of the structural parameters of SFGs and SMGs in four redshift bins.

\subsubsection{Effective radii}
\label{sec:Re}

\begin{figure}
    \begin{center}
    \includegraphics[width=0.48\textwidth]{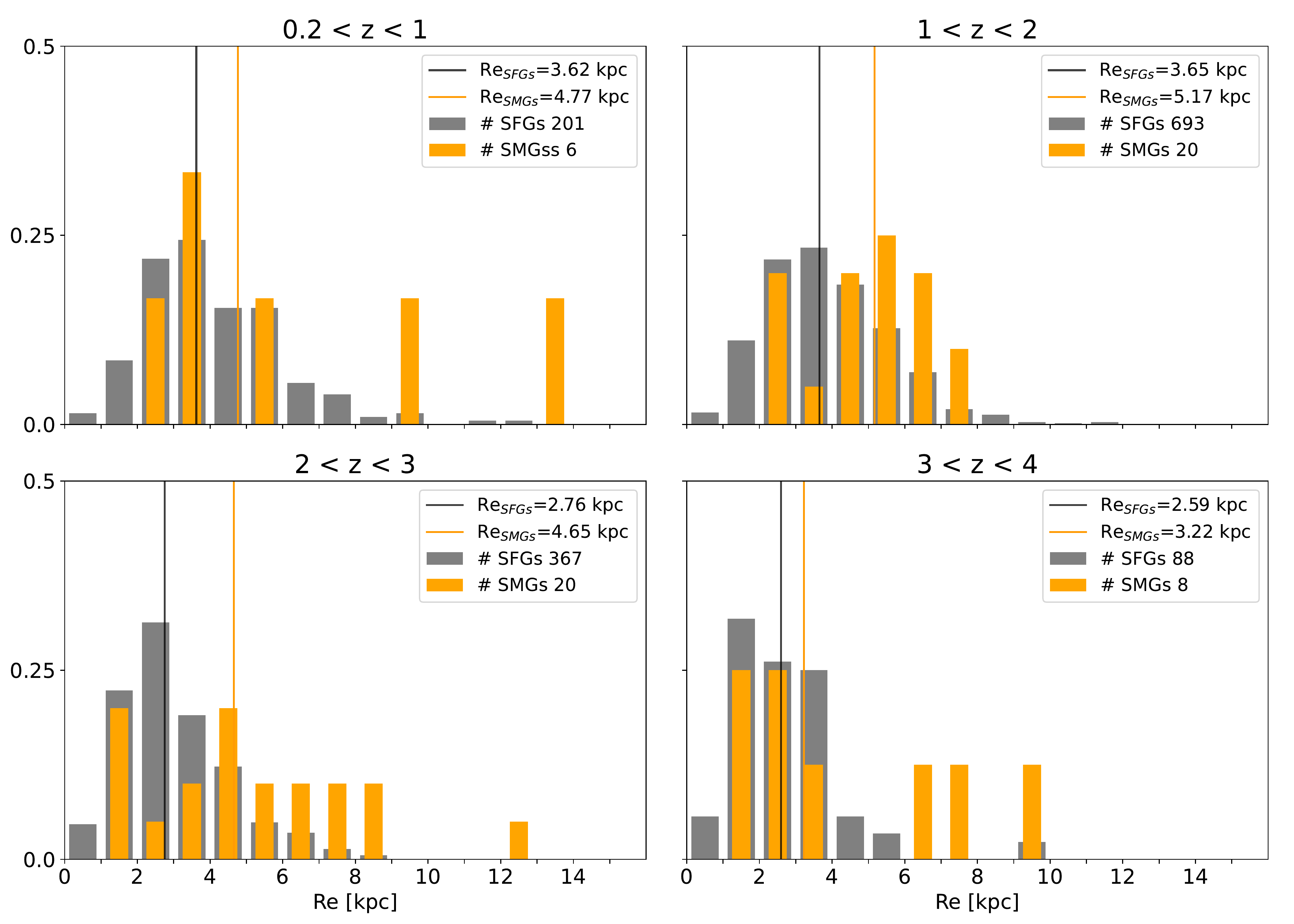}
    \caption{Normalized distribution of the effective radius $R_{\rm e}$ of SMGs (orange) and SFGs (grey) with $\log (M_{\star}/ \rm M_{\odot}) >10$, using $R_{\rm e}$ values estimated by \citet{vdWel2014} and \citet{TesisKarla}. The median effective radii $R_{\rm e}$ of SMGs and SFGs are represented by vertical lines. The median values of the effective radii of SMGs are 50~per cent larger than those of SFGs at $z<3$.}
    \label{fig:hist_Re}
    \end{center}
\end{figure}

\begin{table*}
    \begin{center}
    \caption{Median values of structural parameters for SFGs and SMGs with $\log (M_{\star}/ \rm M_{\odot}) >10$. The columns present: (1) redshift range; (2) median effective radii of SFGs along the semi-major axis; (3) median effective radii of SMGs along the semi-major axis; (4) median radius difference to the size-mass relation of SMGs; (5) median S\'ersic index of SFGs; (6) median S\'ersic index of SMGs; (7) median axis ratio of SFGs and (8) median axis ratio of SMGs. }
    \label{tab:med_params}
    \begin{tabular}{cccccccc}
        \hline
        Redshift & $R_{\rm{e,SFGs}}$ & $R_{\rm{e,SMGs}}$ & $\Delta \log R_{\rm e,SMGs}$ & $n_{\rm{SFGs}}$ & $n_{\rm{SMGs}}$ & $q_{\rm SFGs}$ & $q_{\rm SMGs}$\\
        $z$ & [kpc] & [kpc]& & & & & \\ 
        \hline
        0.2--1.0 & $3.62\pm0.13$ & $4.8\pm0.9$ & $0.09\pm0.09$ & $1.16\pm0.04$ & $1.7\pm0.4$  & $0.52\pm0.02$ & $0.64\pm0.17$\\ 
        1.0--2.0 & $3.65\pm0.10$ & $5.2\pm0.4$ & $0.006\pm0.03$ &  $1.03\pm0.02$ & $1.2\pm0.3$ & $0.55\pm0.01$ & $0.67\pm0.06$\\
        2.0--3.0 & $2.76\pm0.09$ & $4.7\pm0.6$ & $0.22\pm0.05$ &  $1.19\pm0.08$ & $1.2\pm0.22$ & $0.55\pm0.02$ & $0.47\pm0.07$\\
        3.0--4.0 & $2.59\pm0.19$ & $2.5\pm1.6$ & $-0.05\pm0.07$ &  $1.23\pm0.11$ & $2.4\pm1.0$ & $0.51\pm0.03$ & $0.68\pm0.11$ \\ 
        \hline
        Median & $3.33\pm0.05$ & $4.8\pm0.4$ & $0.11\pm0.07$ &  $1.11\pm 0.02$ & $1.23\pm0.22$ & $0.550\pm0.008$ & $0.62\pm0.05$\\ 
        \hline
    \end{tabular}
    \end{center}
\end{table*}

The optical counterparts of faint SMGs have a median effective radius along the semi-major axis $R_{\rm e, SMG}=4.8\pm0.4~\rm kpc$, larger than the median effective radius of the SFG sample $R_{\rm e,SFG}=2.71\pm0.03~\rm kpc$. Since we had found that SMGs are more massive than SFGs (section~\ref{sec:analysis_stellarmass}), this result is not entirely surprising.

In Figure~\ref{fig:hist_Re} we show the normalized distributions of sizes for galaxies within four redshift bins. The corresponding medians for galaxies with $\log (M_{\star}/ \rm M_\odot) >10$ are listed in Table~\ref{tab:med_params}. The median $R_{\rm e,SFG}$ is slightly smaller at redshifts $z>2$ than at $z<2$. The same effect is seen in the SMG population, where the median effective radius of SMGs at high redshift ($z>3$) is smaller than that at $z<3$. The sizes of both populations are only comparable at the highest redshift bin. The median $R_{\rm e,SMGs}$ are significantly larger than the median effective radii of SFGs at $1<z<3$ ($p<0.05$, see Table~\ref{tab:pvalues}). 

\begin{figure}
    \centering
    \includegraphics[width=0.48\textwidth]{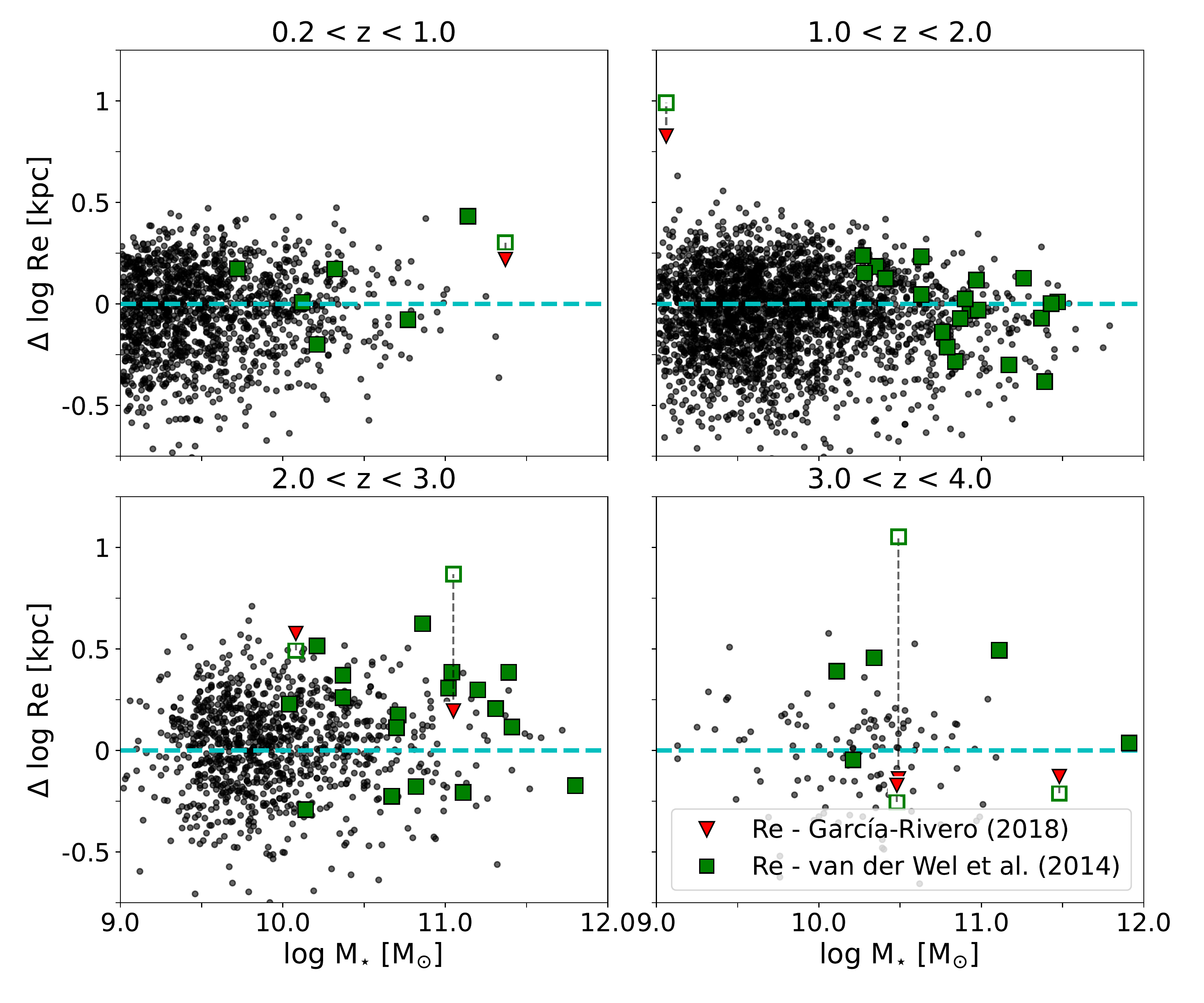}
    \caption{Radii differences vs. stellar mass of the SFGs (black dots) and SMGs (squares and triangles) to the $\log R_{e}-\log M_{\star}$ relation traced by SFGs (cyan dashed line). We represent with green squares the SMGs with $R_{\rm e}$ derived by \citet{vdWel2014}. The empty squares connected to red triangles are those radii replaced with $R_{\rm e}$ derived by \citet{TesisKarla}. The redshift bin where the median size of SMGs is significantly larger than that of SFGs of the same stellar mass is $2<z<3$. }
    \label{fig:logRe_M}
\end{figure}

In order to further explore the sizes of SMGs with respect to the typical sizes of SFGs with the same stellar mass, we fitted a linear $\log R_{\rm e}-\log ~M_{\star}$ relation, following a similar procedure to that employed to derive the main sequence of star-forming galaxies (section~\ref{sec:SFMS}). We then calculate the residual radii, $\Delta \log R_{\rm e}$, as the difference between the galaxy effective radius and the mean effective radius of the SFG sample at the galaxy's stellar mass. In Figure~\ref{fig:logRe_M} we present the residual radii versus stellar mass for SFGs and SMGs and the median $\Delta \log R_{\rm e,SMG}$ values are listed in Table~\ref{tab:med_params}. 

The median effective radii of SMGs are overall significantly larger than the median radii of SFGs of the same mass ($p=0.001$, see Table~\ref{tab:pvalues}). The differences are mainly carried by the population of SMGs at $2<z<3$: probability $p=0.0004$ that the median radii for the same stellar mass could originate from a common distribution. A KS test with $p<0.05$ also confirms that the residual distributions of SMGs and SFGs are different at this redshift bin. At $0.2<z<2$, however, we cannot reject the null hypothesis.

Hence, SMGs are in general larger than the SFG population of the same stellar mass. This difference is more significant at $2<z<3$, where the fraction of starbursts is also larger.

\subsubsection{S\'ersic index}
\label{sec:n}

\begin{figure}
    \centering
    \includegraphics[width=0.48\textwidth]{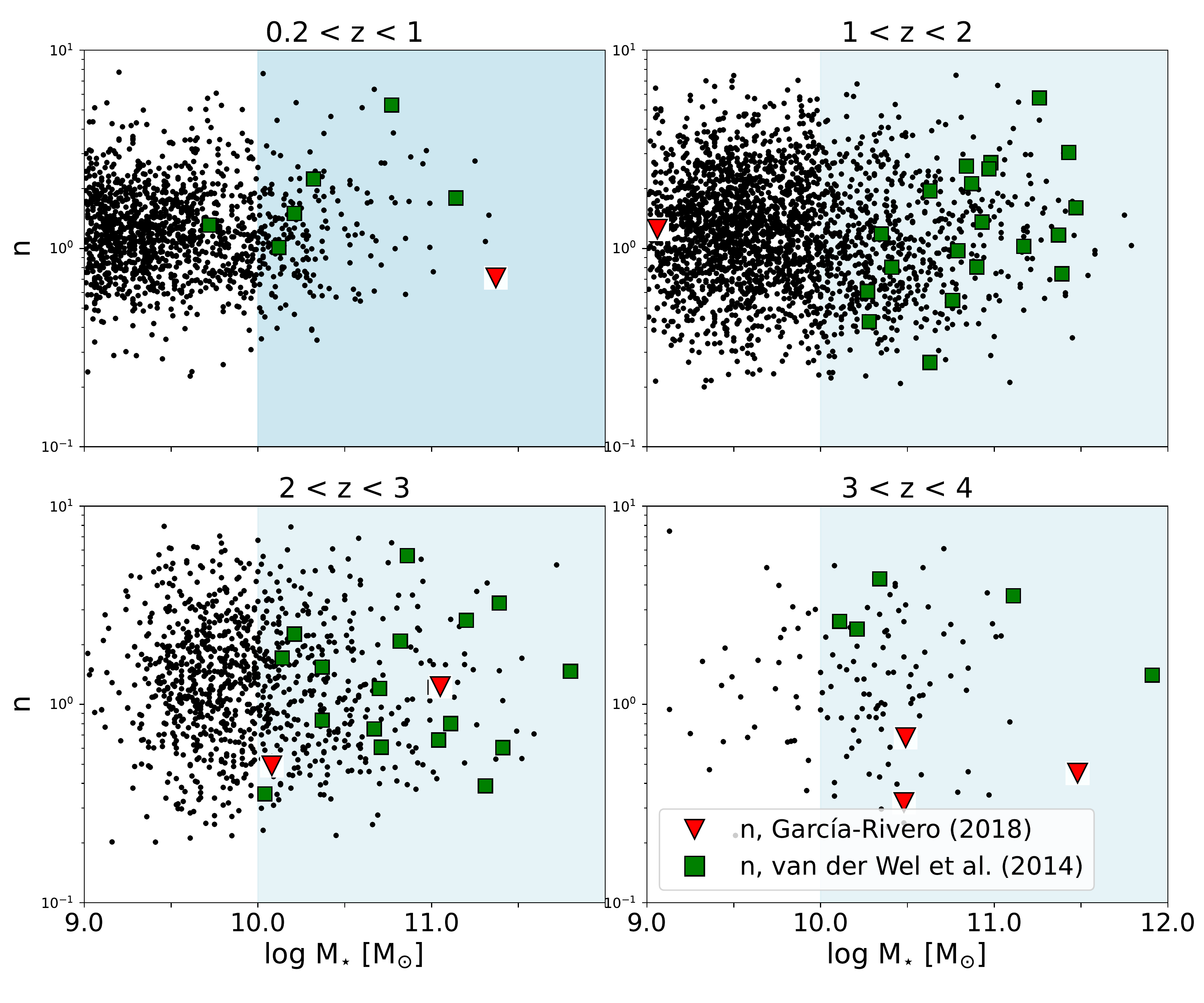}
    \caption{S\'ersic index $n$ vs. $M_{\star}$ for SFGs (black dots) and SMGs (green squares and red triangles), following the same symbols as in Figure~\ref{fig:logRe_M}.
    When we select $\log (M_{\star}/\rm M_{\odot})>10$ galaxies, the indices of SMGs seem to be derived from the same distribution as those of SFGs. }
    \label{fig:n_M}
\end{figure}

\begin{figure}
    \includegraphics[width=0.45\textwidth]{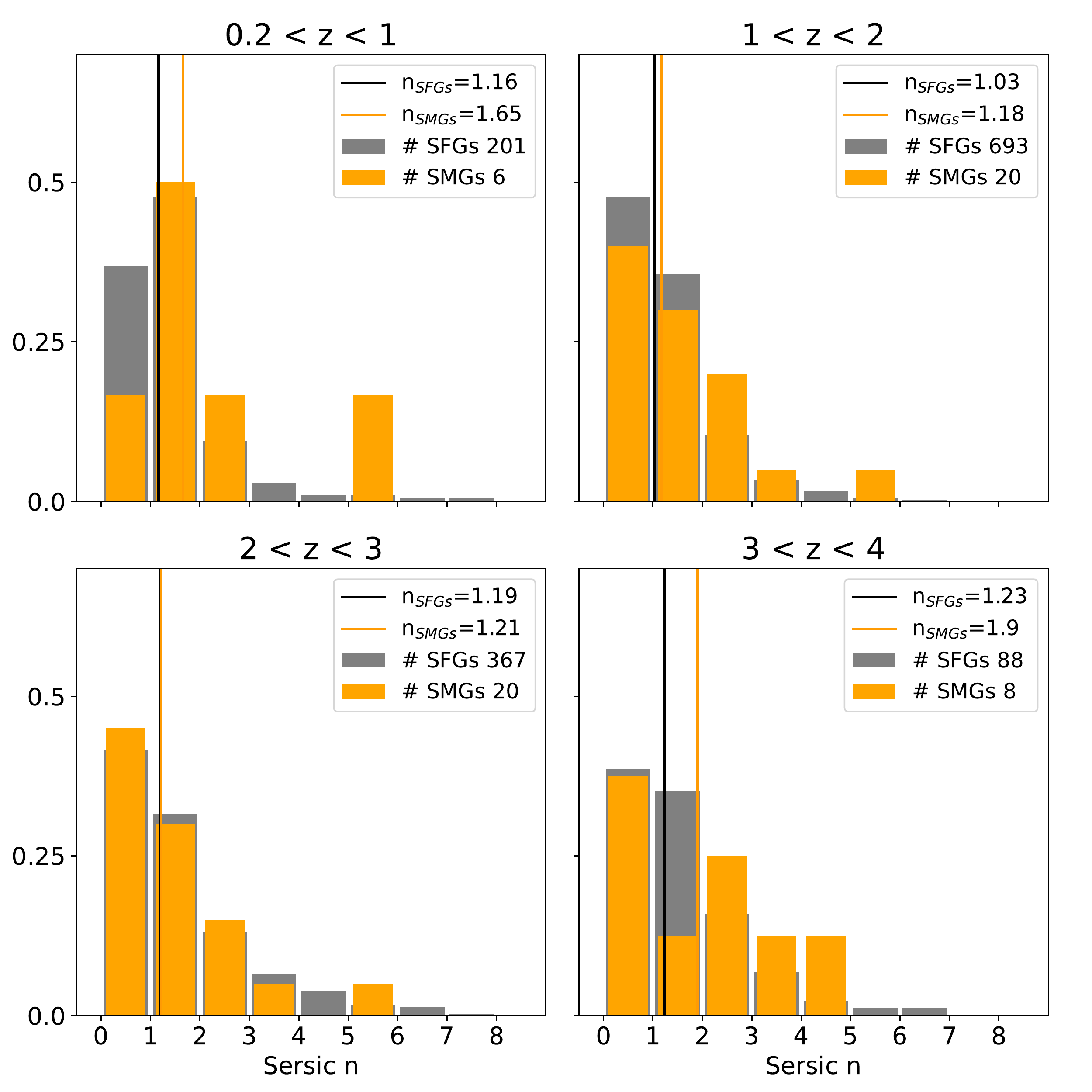}
    \caption{Distribution of the S\'ersic index ($n$) of SMGs (orange) and SFGs (grey) with $\log(M_{\star}/M_{\sun})>10$. Both distributions are normalized by the number of galaxies in each sample. The Mann-Whitney and KS test show that the medians are not significantly different at any redshift.}
    \label{fig:hist_n}
\end{figure}

We present the S\'ersic index vs stellar mass for SFGs and SMGs in Figure~\ref{fig:n_M}. Figure~\ref{fig:hist_n} shows the normalized distribution of the S\'ersic index of both populations with masses $\log (M_{\star}/\rm M_{\odot})>10$ separated in four redshift bins and their corresponding median values are presented in Table~\ref{tab:med_params}. We find no differences between the distributions of $n$ values (Table~\ref{tab:pvalues}).

\citet{Zavala2018} claimed an evolution in the S\'ersic index of SMGs between redshift bins  $0.2<z<1.4$ and $1.4<z<3$, such that $n$ increases towards lower redshifts, which we also confirm. Once we split the sample in four redshift bins, however, the evolution is not as clear. The Mann-Whitney test indicates that the median S\'ersic indices of SMGs and SFGs in this mass range are compatible with a common parent distribution ($p=0.3$) at  $1.4<z<3$.  At $0.2<z<1.4$, however, the null hypothesis of identity is rejected and the median S\'ersic indices are not compatible with a common parent distribution ($p=0.009$): the median S\'ersic index of SMGs at $0.2<z<1.4$ is larger than that of SFGs of the same stellar mass. We will test this result further in section~\ref{CAS}.

\subsubsection{Axis ratio} 
\label{sec:q}

The projected axis ratio $q=b/a$, describes the \textit{roundness} or elongation of the galaxy, and varies from 0 (very elongated projected shape) to 1 (circular). The median values of axis ratios for SMGs and SFGs in each redshift bin are listed in Table~\ref{tab:med_params}. The median axis ratio for SMGs is $q_{\rm SMGs}=0.62\pm0.05$ and for the optically-selected SFGs in the comparison sample is $q_{\rm SFGs}=0.550\pm0.008$. This implies that the projected shape of SMGs is slightly rounder and the Mann-Whitney test shows this difference is statistically significant ($p=0.04$, see Table~\ref{tab:pvalues}). 

\begin{figure}
    \includegraphics[width=0.45\textwidth,height=0.3\textheight]{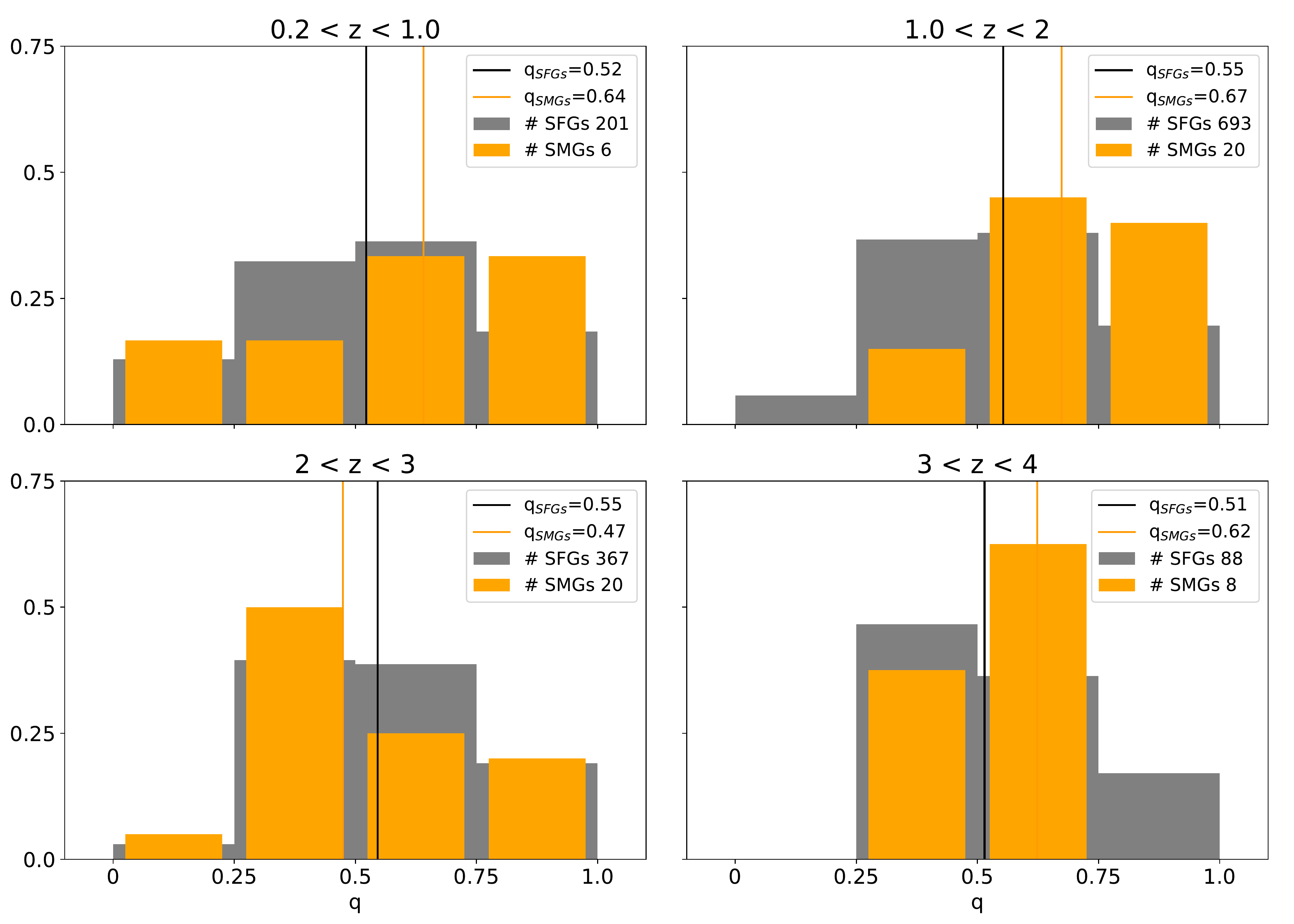}
    \caption{Normalized distribution of the axis ratio ($q=b/a$) for SMGs (orange) and SFGs (grey) with $\log(M_{\star}/{\rm M}_{\sun})>10$. The median $q$ (vertical lines) of SMGs are larger at all redshifts, except $2<z<3$. The Mann-Whitney test indicates that this difference is significant only at $1<z<2$.}
    \label{fig:hist_q}
\end{figure}

We present the normalized distributions of axis ratios in four redshift bins in Figure~\ref{fig:hist_q}. The median $q_{\rm SMG}$ are slightly larger than $q_{\rm SFG}$ at all redshifts, except $2<z<3$, where they have similar values. We applied the Mann-Whitney test to evaluate if the medians stem from the same parent distribution, and find the difference could be significant only at $1<z<2$ ($p=0.01$).

\subsubsection{CAS parameters}
\label{CAS}

The morphology of galaxies can also be described with non-parametric indices like concentration (C), asymmetry (A) and clumpiness (S) \citep{CAS_Conselice2003, Gini-M20_Lotz2004}. Since these indices do not assume a specific function of the light distribution, they are specially useful as we move to higher redshifts or explore irregular and merger populations. We calculated these indices for the SFG and faint SMG samples using the same procedure as in \citet{Gini-M20_Lotz2004}, taking special care of the calculations for the SMG counterparts in the procedure, as a good fraction of these are faint and do not have available CAS indices in the CANDELS catalogs.

\begin{figure}
    \centering
    \includegraphics[width=0.45\textwidth]{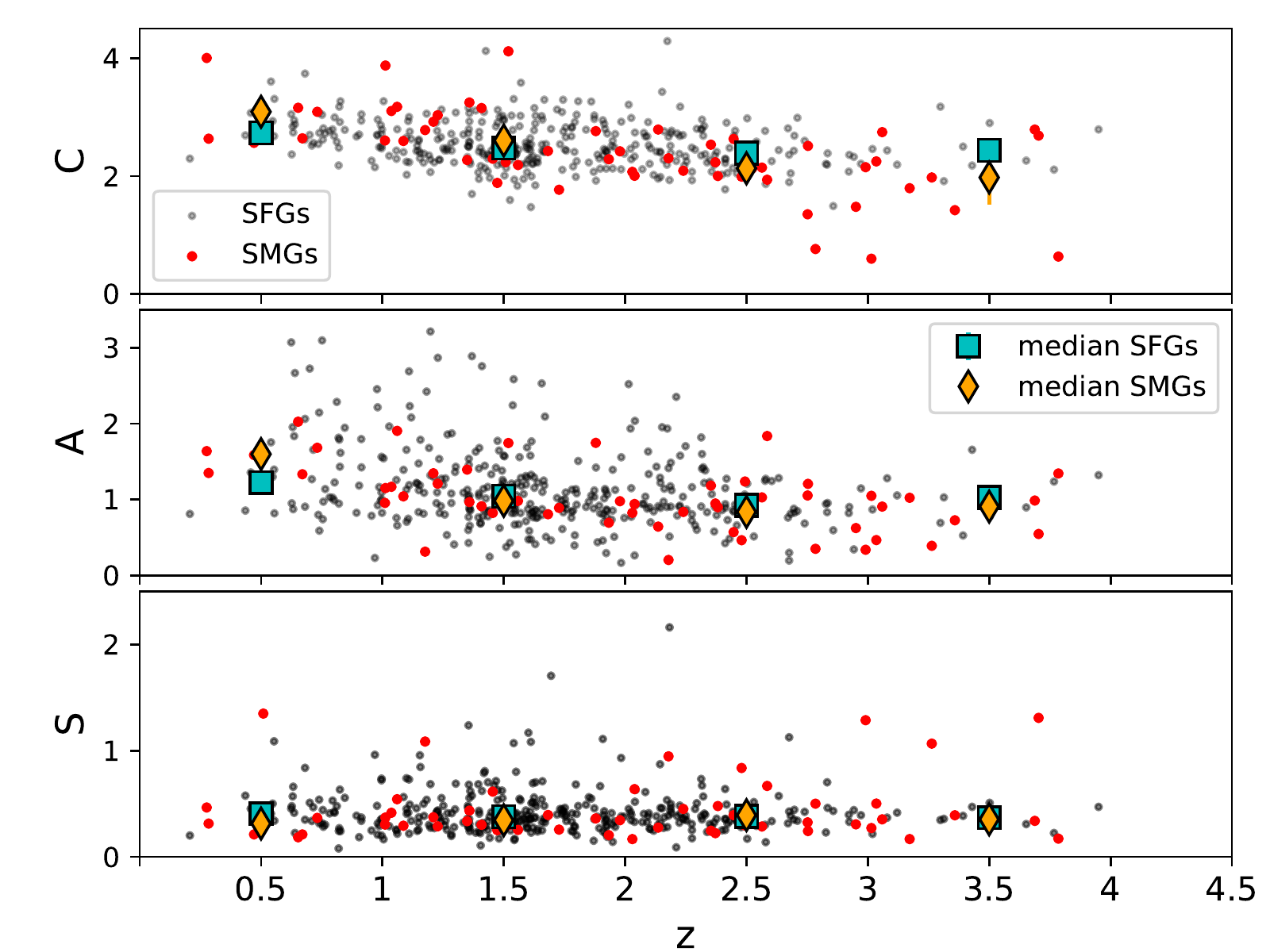}
    \caption{Non-parametric morphology indices for SFGs (black dots) and faint SMGs (red dots) with $\log (M_{\star}/\rm M_{\odot})>10$: (\textit{Top}) Concentration (C), (\textit{Middle}) asymmetry (A) and (\textit{Bottom}) clumpiness (S). The mean of the SFG indices are marked by blue squares in four redshift bins, and the mean of the SMG indices with orange diamonds. We find a correlation between redshift and the concentration and asymmetry for SFGs and SMGs, and differences in median concentrations, such that SMGs have higher concentrations at lower redshift.}
    \label{fig:CAS}
\end{figure}

In Figure~\ref{fig:CAS} we present the CAS indices versus redshift for SFGs and SMGs with $\log (M_{\star}/\rm M_{\odot})>10$. At low $z$ the C and A indices show higher values for both samples, implying an apparent higher concentration and asymmetry of the galaxies in \textit{H}-band.

We explore the correlation between the indices and redshift with the Kendall rank correlation index. We find a strong correlation for the SFGs between the C index and $z$ ($p=5.5\times 10^{-7}$) and A and $z$ ($p=9\times 10^{-7}$). The SMGs show a correlation as well ($p=6\times10^{-7}$ for C vs. $z$ and $p=4\times10^{-4}$ for A vs. $z$). Both SMGs and SFGs show no trend between the clumpiness index of the galaxies and redshift ($p\sim0.5$). 

The median values of the CAS indices can be found  in Table~\ref{tab:CAS_medians}. The differences in the medians of the concentration parameter C are significant globally (see Table~\ref{tab:CAS_pvalues}). The difference is most significant at $2<z<3$ ($p=9\times 10^{-4}$) and marginal, when uncertainties in redshift are taken into account, at $0.2<z<1$ ($p\sim0.05$). The differences in A and S between both populations are not significant.

We hence find a larger concentration for both SMGs and SFGs with decreasing redshift, and the growth in concentration to be larger for SMGs than for SFGs at later times, confirming the results we found for the S\'ersic $n$ index in section~\ref{sec:n}.

\begin{table*}
    \centering
    \caption{Median values of the non-parametric indices for SFGs and SMGs with $\log (M_{\star}/ \rm M_{\odot}) >10$. The columns present: (1) redshift range; (2) median concentration index of SFGs; (3) median concentration index of SMGs; (4) median asymmetry of SFGs; (5) median asymmetry of SMGs; (6) median clumpiness index of SFGs and (7) median clumpiness index of SMGs.}
    \begin{tabular}{c|c|c|c|c|c|c}
        Redshift & C$_{\rm SFGs}$ & C$_{\rm SMGs}$ & A$_{\rm SFGs}$ & A$_{\rm SMGs}$ & S$_{\rm SFGs}$ & S$_{\rm SMGs}$\\ \hline
        0.2-1.0 & $2.73\pm0.03$ & $3.11_{-0.003}^{+0.04}$ & $1.23_{-0.11}^{+0.14}$ & $1.62\pm0.04$ & $0.41\pm0.04$ & $0.29\pm0.11$ \\
        1.0-2.0 & $2.47\pm0.03$ & $2.60\pm0.18$ & $1.04\pm0.03$ & $0.98\pm0.05$ & $0.38\pm0.01$ & $0.35\pm0.03$ \\
        2.0-3.0 & $2.39\pm0.03$ & $2.13\pm0.09$ & $0.92\pm0.02$ & $0.84\pm0.07$ & $0.38\pm0.02$ & $0.39\pm0.06$ \\
        3.0-4.0 & $2.43\pm0.25$ & $1.98^{+0.28}_{-0.24}$ & $1.03\pm0.19$ & $0.91^{+0.13}_{-0.10}$ & $0.37\pm0.02$ & $0.35\pm0.06$ \\ \hline
        $0.2-4.0$ & $2.49\pm0.02$ & $2.34\pm0.11$ & $1.03\pm0.03$ & $0.98\pm0.03$ & $0.38\pm0.01$ & $0.36\pm0.02$ \\
         \hline
    \end{tabular}
    \label{tab:CAS_medians}
\end{table*}

\subsubsection{Machine-learned classification}
\label{sec:visual-morph_fractions}

\begin{figure}
    \centering
    \includegraphics[width=0.48\textwidth]{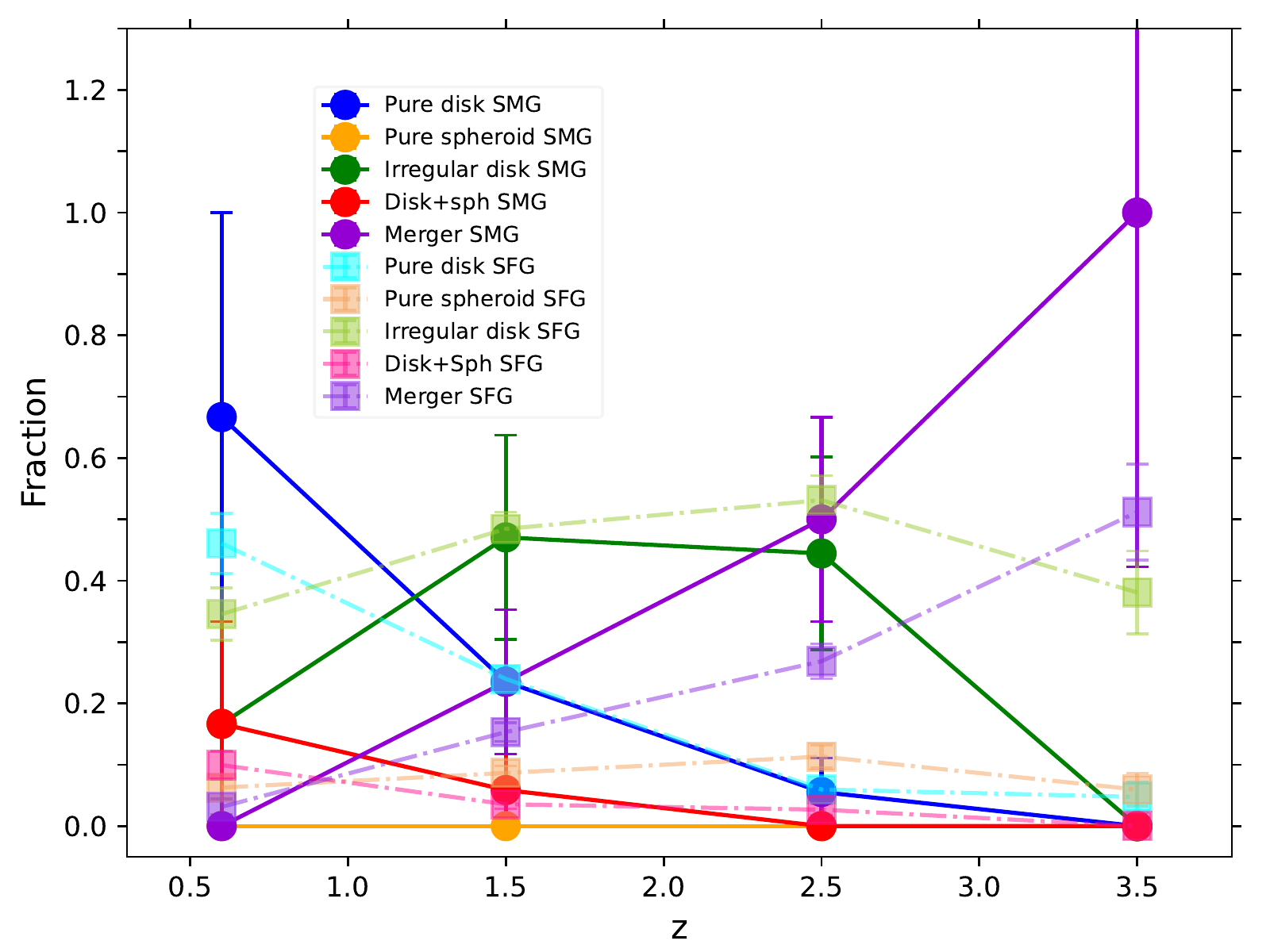}
    \caption{Fraction of SMGs (circles) and SFGs (squares) with $\log(M_{\star}/M_{\sun})>10$ that are classified as pure disks, pure spheroids, disks+spheroids, irregular disks and irregulars/mergers, from the machine learning visual classification catalog. }
    \label{fig:vismorph_fractions}
\end{figure}

In order to further check possible discrepancies between the SFG and faint SMG populations, we use the morphological classes presented by \citet{HuertasCompany}: pure disks, pure spheroids, disks+spheroids, irregular disks and irregular/mergers. We applied their criteria to both SMGs and SFGs with $\log (M_{\star}/ \rm M_{\odot}) >10$, finding classifications for 44 SMGs and 975 SFGs. In Figure~\ref{fig:vismorph_fractions} we present the fraction of SMGs and SFGs classified by morphological type at each redshift bin.

We find that the predominant morphological type for faint SMGs is disk-like galaxies at $z<2$ (pure disks, irregular disks and disks+spheroids), while at $z>2$ the fraction of mergers is roughly the same as that of irregular disks. All SMGs and half of SFGs (51~per cent) at the highest redshift bin $3<z<4$ are classified as mergers. However, there are only 3 SMGs with classifications in the highest redshift bin. 

Irregular disks and mergers in the SMG population decrease to give rise to pure disks and disk+spheroids. At the lowest redshift bin 67~per cent of SMGs are classified as pure disks, with equal fractions ($\sim 17$~per cent) of irregular disks and disks+spheroids. SFGs show similar fractions of pure disks and irregular disks in this redshift bin with a $\sim15$~per cent of disk+spheroids. Both galaxy populations show an increment of the disk+spheroid and pure disk fraction towards lower redshifts.

There are no SMGs classified as pure spheroids and there are $\leq 10~$per cent of spheroidal SFGs at any redshift. Overall the main morphology of SMGs is disks, and an evolution in both faint SMG and SFG populations can be seen, where the merger fraction decreases with redshift and, irregular disks dominate at intermedate redshifts ($1<z<3$), while pure disks and disk+spheroids rise at the lowest redshift bin. There are no clear differences in the evolution of both populations.

The selection criteria we use in this work is that of \citet{HuertasCompany}, which require probabilities $>2/3$ for disks and spheroids classification, and produces a different result to that applied by \citet{Zavala2018}, where probabilities $>1/3$ were used. This difference results in a smaller number of galaxies classified as pure spheroids and disk+spheroids.

In section~\ref{sec:SMGlocMS} we found that 82~per cent of SMGs are located above the main sequence of star formation, and 40~per cent can be classified as starbursts. Among starbursts galaxies we find 27~per cent to have merger morphologies and 73~per cent to have disk-like morphologies: 61~per cent irregular disks, 6~per cent disk+spheroids and 6~per cent pure disks. On the other hand, the morphologies of the faint SMGs that lie within $\pm 1\sigma$ of the main sequence are classified as  24~per cent mergers, 35~per cent irregular disks, 35~per cent pure disks and 6~per cent disk+spheroids. Hence there are slightly more starburst SMGs classified as mergers and irregular galaxies than among main-sequence faint SMGs, but the differences are within the poisson errors of the samples. However, at the redshift bin where we find significantly larger sizes for SMGs than for SFGs and a larger number of starbursts ($2<z<3$, section~\ref{sec:Re}), the galaxies above the $\log R_{\rm e}-\log M_{\star}$ sequence are classified as mergers (43~per cent) and irregular disks (57~per cent). This could be a trace of recent interactions.  

\subsection{The impact of possible misidentifications of counterparts of faint SMGs}

Our analysis is based on the identifications of optical-infrared counterparts presented in \citet{Zavala2018}. There is the potential of misidentification, which was estimated to be of 13~per cent for the full sample. To miminize the impact of misidentifications, six sources with discrepant photometric redshifts from optical-IR and FIR-radio methodologies were discarded. 

We nevertheless adopt a 13~per cent contamination into our final sample of 57 SMGs and explore if the conclusions derived in sections~4.1 to 4.3 are robust regardless of the possible contamination of misidentified counterparts. In order to do this, we randomly assign to 13~per cent of the faint SMGs the properties of another SMG in the sample with similar FIR-colours (indicative of redshift), irrespective of their brightness. We recomputed all estimations for median stellar mass, star formation rate, size and morphology per redshift bin. We find that the results do not change and our conclusions for the mean properties of the population are robust.

\section{Discussion}
\label{sec:discussion}

\subsection{Stellar Mass} 
\label{sec:discussion_stellarmass}

In section~\ref{sec:analysis_stellarmass} we found that the stellar masses of SMGs are consistent with an average value of $\log (M_{\star}/\rm{M_{\odot}})=10.75\pm0.07$ across all redshifts. We checked that dust obscuration as measured by $A_V$ was not biasing this result. We also showed, however, that $A_V$ was not able to correct for the total obscuration of SMGs alone (section~\ref{sec:SFRcompa}). Hence, the question still remains whether the stellar mass enshrouded by dust associated to the bulk of $L_{\rm IR}$ emission is significant compared to the mass measured through the most complete optical-MIR photometry available.

\citet{Michalowski2014} simulated a sample of SMGs and their synthetic photometry, finding that a double-peaked burst was the best Star Formation History (SFH) to reproduce the measured colors of SMGs. An exponentially declining SFH returns stellar masses that are lower, but still consistent with their inputs. They also found that the correct age plays a more important role in the estimation of the stellar mass than dust extinction $A_{V}$. Therefore, even though  $\rm{SFR_{\rm UV}^{corr}}$ cannot account for the total SFR of SMGs, the masses, which depend on the rest-frame infrared flux densities, are not heavily affected. Based on their results, we estimate that the stellar masses of SMGs could be underestimated by 0.3-0.5 dex, which is a factor of $\sim 2$ times the intrinsic scatter between methods found in the CANDELS catalogs \citep{Mobasher}. 

We tested the effects of this possible bias, considering a stellar mass 0.3 and 0.5 dex larger than the original stellar mass reported in the CANDELS catalog. The sSFRs are consequentely 0.3 and 0.5 dex smaller. The bias hence would  displace SMGs in a diagonal line in Figure~\ref{fig:SFMS} towards higher $M_{\star}$ and lower sSFR.

The Mann-Whitney results for the $\Delta \log \rm sSFR$ medians of SFGs and SMGs remain the same for a 0.3~dex stellar mass offset. We would find in this situation 71~per cent of faint SMGs above the main sequence (as compared to the original 82~per cent), and 29~per cent of starbursts (instead of 40~per cent). If we adopt a 0.5~dex stellar mass offset, we would find 61~per cent of faint SMGs above the main sequence and 16~per cent of starbursts. The highest fraction of SMGs above the main sequence still is at $2.5<z<3$.

Hence, considering the possible stellar mass bias, we would find more faint SMGs tracing the main sequence and fewer starbursts, while a large fraction of faint SMGs still populate the upper parts of the scatter of the main sequence.

\subsection{Starbursts among faint SMGs?}

In this paper we derive that 35--40~per cent of faint SMGs are starbursts, based on the $R_{\rm SB}=\rm{SFR/SFR_{MS}}>3$ criterion and our best estimates of stellar mass and SFR (sections~\ref{sec:SMGlocMS} and ~\ref{sec:SFRcompa}).

In previous works the fraction of SMGs classified as starburst varies according to the selection wavelength and depth of the catalog. 

\citet{Zavala2018} reported that 85~per cent of their sample of 72 faint SMGs lie within the $3\sigma$ scatter of the main sequence of \citet{Speagle2014}. After the exclusion of the SMGs without CANDELS and 3D-HST counterparts and  discriminating by redshift bins, we find a 96~per cent fraction of faint SMGs within the $3\sigma$ scatter of the main sequence we derived. We note that the main sequence adopted in \citet{Zavala2018} is shallower than that derived in this paper at $z>2$, and their SFR slightly smaller than those adopted here, and hence the location of SMGs with respect to the main sequence is different. We also find that the residual sSFR to the main sequence for these faint SMGs is significantly different to the distribution of SFGs at $z>1$, even when possible stellar mass biases are considered (section~\ref{sec:discussion_stellarmass} ).

\citet{DaCunha_2015} followed-up with ALMA a sample of SMGs with $S_{870 \mu\rm m} > 4~\rm mJy$. They found that half of their SMGs are located above the main sequence with $R_{\rm SB}>3$. We find similar fractions of galaxies above the main sequence in our sample of faint SMGs, when we derive the fractions at similar depths.

In a 1.3 mm ALMA survey of 16 bright galaxies with $S_{1.1\rm mm} > 3.3~\rm mJy$, \citet{Miettinen2017aztec} estimated 63~per cent of SMGs are starbursts with $R_{\rm SB}>3$. They hence found a much higher fraction of starburst galaxies, which could be due to the selection wavelength and brightness of the sample. In a subsequent study, \citet{Miettinen2017} imaged a larger sample of the 129 brightest AzTEC sources in the COSMOS field with ALMA at 1.3 mm, finding 57~per cent of galaxies within the main sequence scatter (and below $R_{\rm SB}\sim3$). From the starburst SMGs 49~per cent are preferentially located at $z>3$. This differs from our finding of the highest fraction of starbursts being located at $2.5<z<3$, but our statistics at the highest redshift bin are poor. 

\citet{Franco20} studied a sample of 35 ALMA 1.1 mm-detected galaxies in the GOODS-S field, finding that 54~per cent of them have an offset to the main sequence over $R_{\rm SB}>3$. The catalogue is created from a sample of 19 sources detected in a blind survey  at  $S_{1.1 \rm mm} \geq 0.7~ \mu\rm Jy$ and 16 additional galaxies detected at lower S/N using the \textit{Spitzer}/IRAC and VLA counterparts. Therefore, they reached a deeper selection limit due to the supplementary catalogue, but found a fraction of starbursts consistent with other samples selected at the same wavelength. 

\citet{STUDIES-III_450_Lim2020} studied a sample of $450~\mu\rm m$-selected faint SMGs in the STUDIES survey with $L_{\rm IR}\sim 10^{11}~{\rm L_\odot}$. They found that 35~per cent of the SMGs are starburst.

When the different selection wavelengths, depths, definition of star-formation main sequence and the small sizes of the samples are taken into account, we see an overall agreement in the \textit{starburstiness} of faint SMGs of around 30-50~per cent, while classical SMGs have reported starbustiness in excess of 50~per cent.

\subsection{Star formation rate indicators of faint SMGs and patchy obscuration}

In section~\ref{sec:SFR} we found that the absence of mid-to-far IR data heavily impacts the determination of the total SFR of faint SMGs due to an extrapolation from the rest-frame UV to NIR estimation of obscuration. Counter-intuitively, we found that sources with the lowest optically-inferred obscuration ($A_V$) were also those that had the strongest underestimations of their total SFR, based on $\rm{SFR_{UV}^{corr}}$. This dust extinction was estimated using the UV continuum's slope, as well as the IRX-$\beta_{UV}$ relation to correct (see section~\ref{sec:data_source}). However, as it was indicated in section~\ref{sec:SFRcompa}, there are discrepancies between the CANDELS and S2CLS teams estimation of $L_{\rm IR}$. Our interpretation is that this is the direct result of patchiness in the dust distribution within these galaxies, such that the estimations from UV-optical data come from different regions than the FIR emitting ones. 

A similar conclusion was derived by \citet{Wuyts2011_SFRs}, who found that dust-correction methods to infer the SFR of the highest star-forming galaxies ($\rm{SFR>100~M_{\odot}~yr^{-1}}$, especially at $z>2.5$) fail to recover the total amount of star formation, when compared to the $\rm{SFR_{FIR}}$ derived from {\it Herschel} $70-160~\mu\rm m$ data.  They estimate that the templates enhanced with PAH emission are not enough to reproduce the FIR contribution to the total SFR from $24~\mu\rm m$ fluxes for these high star-forming systems, which are affected by patchy dust obscuration. The uneven dust obscuration causes the saturation of the UV-slope derived extinction, $A_{V}$. This effect has also been seen in local LIRGs and starburst galaxies that are located above the $\rm IRX-\beta$ relation \citep{MeurerIRX-beta, Goldader2002, Howell2010}. Some of our SMGs are indeed located above the IRX-$\beta$ relation \citep{Zavala2018}. This effect could be due to various factors, including differing star-formation histories \citep{Salmon_2016, Calzetti_revisit-attcurves}. 

\citet{Elbaz2018} observed a sample of massive starburst galaxies selected in \textit{Herschel} bands and a complementary sample of 1.3 mm-selected galaxies. They found that the FIR and the UV emissions of starburst galaxies with $R_{\rm SB}=\rm{SFR/SFR_{MS}}>3$ had systematic spatial offsets. Their SFRs estimated by fitting the UV-to-near IR fluxes is consistent with $\rm{SFR_{UV+IR}}$ for $R_{\rm SB} <1$, but increasingly discrepant above the main sequence. In our sample of SMGs 40~per cent have $R_{\rm SB}>3$. 

This further supports our claim that the optical dust correction underestimates the total SFR for the population of dusty star-forming galaxies.

\subsection{\textit{H}-band Morphology of Faint SMGs}

In section~\ref{sec:visual-morph_fractions} we found that most SMGs have a disk-like morphology according to the machine classification, and in section~\ref{sec:SMGlocMS} we found that 82~per cent of faint SMGs are above the main sequence relationship between sSFR and stellar mass. 
The location of galaxies in the main sequence has been linked to the morphology, where main sequence (U)LIRGs are dominated by non-interacting disks, most galaxies classified as starbursts are either irregular disks or mergers, and spheroids and disk+spheroids are below the main sequence \citep{Kartaltepe15,Osborne2020}. Considering that 32~per cent of the SMGs in our sample lie within $1\sigma$ of the main sequence we find consistently that $\sim76$~per cent are disk-like from visual morphology classifications. We find decreasing fractions of merger SMGs and SFGs towards lower $z$, and lower fractions of merger SFGs than SMGs at all redshifts. There are 35~per cent of mergers in the whole sample of faint SMGs, which is similar to the 40~per cent fraction of starburst found in our sample. Among these starburst SMGs we find indeed 88~per cent of them can be classified as mergers or irregular disks, but we also find these morphologies within the main sequence. 

Based on S\'ersic fits, however, $\sim 95$~per cent of classical SMGs have been found to be best described as massive $\log (M_{\star}/ \rm M_{\odot}) \sim 11.3$ disk-like galaxies, while only $\sim 25$~per cent could be classified as mergers based on the presence of multiple clumps in $H$-band images \citep{Targett2013}. Visual classification on classical SMG samples finds also a morphological evolution with redshift, such that merging and irregular morphologies give way to more ordered disk morphologies at lower reddshifts \citep{LingYang2022}.

We further find in our analysis (sections~ \ref{sec:n} and \ref{CAS}) that both SMGs and SFGs with $\log (M_{\star}/ \rm M_{\odot}) >10$ increase their concentration towards lower redshifts ($n$ and $C$), but they do so at different rates. SMGs are more concentrated at $z<1$, both considering $n$ and $C$, while at $z>2$ SFGs have larger values of the concentration parameters than SMGs. The cross over of concentration between the populations happens at $1<z<2$. At $0.2<z<1.4$ $n$ is significantly larger in SMGs than in SFGs, while at $2<z<3$ $C$ is significantly smaller in SMGs.

Throughout the evolution of both populations, the bulge component is increasing towards lower redshifts (more disk+sph morphologies at lower redshifts). The statistics for SMGs allows to establish that bulge increase at $z<2$ through the machine-assigned categories, but this classification goes in hand with the increment of $n$ and $C$ in the parametric and non-parametric analysis of the images.

The smooth increase in $C$ and $A$ non-parametric indices at a fixed observed band has been found in other samples \citep{Whitney_CAS}, and these are in part attributed to rest-frame pass-band changes and reduced resolution. Since our goal is to make a direct comparison to SFGs at the same redshift, a correction for these effects is not necessary.

We find a significant difference in the sizes of SMGs and SFGs at $z<3$ (section~\ref{sec:Re}), that is mainly driven by the difference in mass selection of the SFG and SMG samples. However, when we study the offset of both populations from the $\log R_{\rm e}-\log M_{\star}$ relationship, we find that SMGs are $\sim 50$~per cent larger than SFGs of the same mass at $2<z<3$, which is the redshift bin with the largest fraction of starbursts and high sSFR galaxies (65~per cent above $1\sigma$ of the Main Sequence). Some of the SMGs in this redshift bin have close companions or disturbed morphology.

When studying massive galaxies with $\log (M_{\star}/ \rm M_{\odot}) >10$, SFGs have on average monotonically increasing sizes towards lower redshifts. SMGs have larger sizes than the SFGs at $z<3$, but not at $z>3$. This could be due to obscuration effects, that only allow the patchy less obscured areas to be revealed in the highest redshift SMGs $H$-band images. In the  $H$-band images the emission traced corresponds to the rest-frame $U$-band ($\sim0.33~\mu\rm m$) at $z\sim4$ and to the rest-frame NIR ($\sim1.4~\mu\rm m$) at $z\sim0.2$, less susceptible to dust obscuration. Indeed, \citet{Chen2022_JWST_morphSMG} found progresively smaller sizes at longer observation wavelengths (considering $HST$, $Spitzer$ and $JWST$ data) in the analysis of the size using curve of growth for a small sample of faint SMGs, which includes 6 of the galaxies studied in this work.

The median effective radii of the SMGs in our sample are consistent within the uncertainties with the \textit{H}-band radii derived in bright SMG samples and other studies of dusty galaxies below the classical SMG regime \citep{Targett2013, Chen2015, Chang2018, STUDIES-IV_Chen-Fatt2020,Franco20}.
Furthermore, the resolved FIR emission of bright SMGs has been found to be more compact than the optical counterparts \citep{Gullberg_2019,Hodge_2019}.

Hence, while we find significant morphological  similarities between faint SMGs and SFGs, like both populations favoring disk-like galaxies, there are significant differences at $2<z<3$, where SMGs are 50~per cent larger than SFGs of similar mass, starbursts are more prominent and also the morphological classification indicates more disturbed morphologies (81~per cent of irregular disks and mergers). 

The newly released data of the JWST allows some clarity on the resolved morphology of these type of sources. A lensed grand-design spiral galaxy discovered by ALMA, located in an overdense region at $z=3$, shows evidence of a minor merger and asymmetry  \citep[][]{Wu2022_JWST_ALMA}. However, a larger sample of SMGs is needed to establish the possible multiple evolutionary tracks of the dusty star-forming population, as evidenced by \citep{ChengCheng2022_JWST_SMG}. We also find that faint SMGs develop (or reveal) a concentration in their $H$-band images at later times than SFGs of the same stellar mass.

\section{Conclusions} 
\label{sec:conclusions}

We studied the physical and morphological properties of 57 faint submillimeter galaxies detected at 450 and $850~\mu\rm m$ in the S2CLS Extended Groth Strip field, with flux densities in the ranges $S_{850 \mu\rm m}=0.7-6~\rm mJy$ and $S_{450 \mu\rm m}=3-17~\rm mJy$, in the flux density regime below classical submillimeter galaxies ($S_{850 \mu\rm m} \lesssim 6~\rm mJy$).
We compare them to a sample of optically-selected star-forming galaxies of similar mass extracted from the same field and analyzed with the same techniques in order to detect if there are differences between the populations. Our main conclusions are:

\begin{itemize}
    \item  Adopting our best estimates of total SFR, faint SMGs are on average located within the $3\sigma$ scatter of the main sequence of star formation, defined within the same field and with the same methodology. About 82~per cent of faint SMGs are located above the main sequence and only 4~per cent above the $3\sigma$ intrinsic scatter of the main sequence, but even within the main sequence, 40~per cent of our faint SMG sample has starburst characteristics, with $R_{\rm SB} (\rm{SFR/SFR_{MS}}) >3$. Faint SMGs have significantly larger sSFRs at $z>1$ than  optically-selected star-forming galaxies of the same mass. 
    
    \item  The sizes of SMGs and SFGs are significantly different at $2<z<3$. We find that, for galaxies  of the same stellar mass ($\log (M_\star/\rm M_\odot) >10$), SMGs are 50~per cent larger than SFGs. In this redshift bin we also find the largest starburst fraction in the SMG sample. Hence we could be witnessing merging processes, also supported by a high fraction of SMG counterparts with morphological classifications falling in the merger (43 per cent) or irregular disk (57 per cent) classes at this redshift bin.
    
    \item We find an evolution in the morphology of SMGs from mergers and irregular disks at high redshift to pure disks and disk+spheroids at low redshift. A similar evolution is seen in high-mass SFGs, although the fraction of mergers in SMGs is slightly larger.
    
    \item We find an evolution of the S\'ersic index of the optical counterparts of SMGs from $n\sim1.0\pm0.2$ at $1.4<z<3$ to $n\sim1.8\pm0.4$ at $0.2<z<1.4$, consistent with the findings of \citet{Zavala2018}. When we compare this evolution with that of SFGs of the same stellar mass, $\log(M_\star/\rm M_\odot)>10$, we find that SMGs are significantly more concentrated at $z<1.4$. The same redshift evolution can be inferred from the non-parametric concentration index C. The concentration of SMGs is significantly smaller than that of SFGs at $2<z<3$ and marginally larger at $0.2<z<1$. No differences can be found with the asymmetry A and clumpiness indices between both populations. 
 
    \item We find that the $\rm SFRs$ estimated without the use of FIR data are underestimated by factors of 3--570 with respect to the FIR-based SFR measurements. The intense obscuration in these systems tends to underestimate the SFR when using only optical-to-near IR data. These systems have the lowest $A_{V}$ values from the whole SMG sample. This is likely an effect of differential obscuration and patchiness at rest-frame UV-to-optical wavelengths. This implies that the optical dust extinction parameter is unable to correct the absorption by dust in heavily obscured systems and, consequently, the total SFR from optical-to-near IR data alone is underestimated. The UV-optical emission could correspond to less obscured regions, implying patchiness or core concentration in the dust distribution, which could be further studied with higher resolution FIR imaging.
    
    \item We find a discrepancy between the FIR-based SFRs derived from different teams, which we trace to the different methodologies to extract flux densities from the \textit{Herschel} data.
    The median of the ratios is $\langle \rm{SFR_{FIR} +SFR_{UV}}/ \rm{SFR_{UV+IR}} \rangle = 0.59\pm0.04$. Although in principle, this discrepancy coud have consequences for the number of faint SMGs above the main sequence and the starbursts in the sample, these numbers are very similar for both SFR estimations.

\end{itemize}

The initial understanding of SMGs located them as high redshift analogues of ULIRGs \citep{Sanders2003}, although detail morphological analysis supported the view that they were mainly massive isolated disks experiencing large star formation rates \citep{Targett2013}. SMGs have also been linked as possible precursors to QSOs and elliptical galaxies \citep{Granato2001}. The faint SMGs in our sample are also hosted in massive disk-like galaxies, while we have shown that they grow spheroidal components more prominently at later times than optically selected SFGs of the same mass. We also find larger effective radii at $2<z<3$, where more starbursts are present. This is also the redshift bin where irregular disks and mergers are dominant morphological classes, possibly reflecting close interactions that increase star-formation, in excess of those found among the optically-selected SFGs. Therefore, this population of SMGs might be the link between the extreme starbursts and main sequence massive star-forming galaxies, represented by the SFG behaviour. 

The future exploration of wider and deeper submillimeter surveys with larger single dish and larger samples of interferometric resolved follow-up will allow better statistical comparisons between galaxy populations. Furthermore, the exploration of the gas content of SMGs will indicate whether these are post-starburst or highly efficient star-forming galaxies.

\section*{Acknowledgements}
We thank the referee for useful comments that significantly improved the sistematics in the conclusions of the paper. This work has been funded by CONACYT grants FDC2016-1848 and CB2016-81948. A.M. thanks support from CONACYT grant A1-S-45680. This work has made use of the Rainbow Cosmological Surveys Database, which is operated by the Centro de Astrobiología (CAB/INTA), partnered with the University of California Observatories at Santa Cruz (UCO/Lick,UCSC).

\section*{Data availability}

The S2CLS catalogue of submillimeter galaxies used in this article is available at \url{https://doi.org/10.1093/mnras/sty217}. The catalogues of the CANDELS program can be found at \url{https://archive.stsci.edu/missions/hlsp/candels/egs/catalogs/v1/}. The morphological parameters catalogue by \citet{vdWel2014} is available at \url{https://www2.mpia-hd.mpg.de/homes/vdwel/3dhstcandels.html}. Multiple catalogues used here can be explored in the Rainbow navigator, at \url{https://arcoirix.cab.inta-csic.es/Rainbow_navigator_public/}.


\bibliographystyle{mnras}
\bibliography{refs} 


\appendix
\section{Color selection for all stellar mass and redshift bins} \label{UVJ_ext}

\begin{figure*}
    \includegraphics[width=\textwidth]{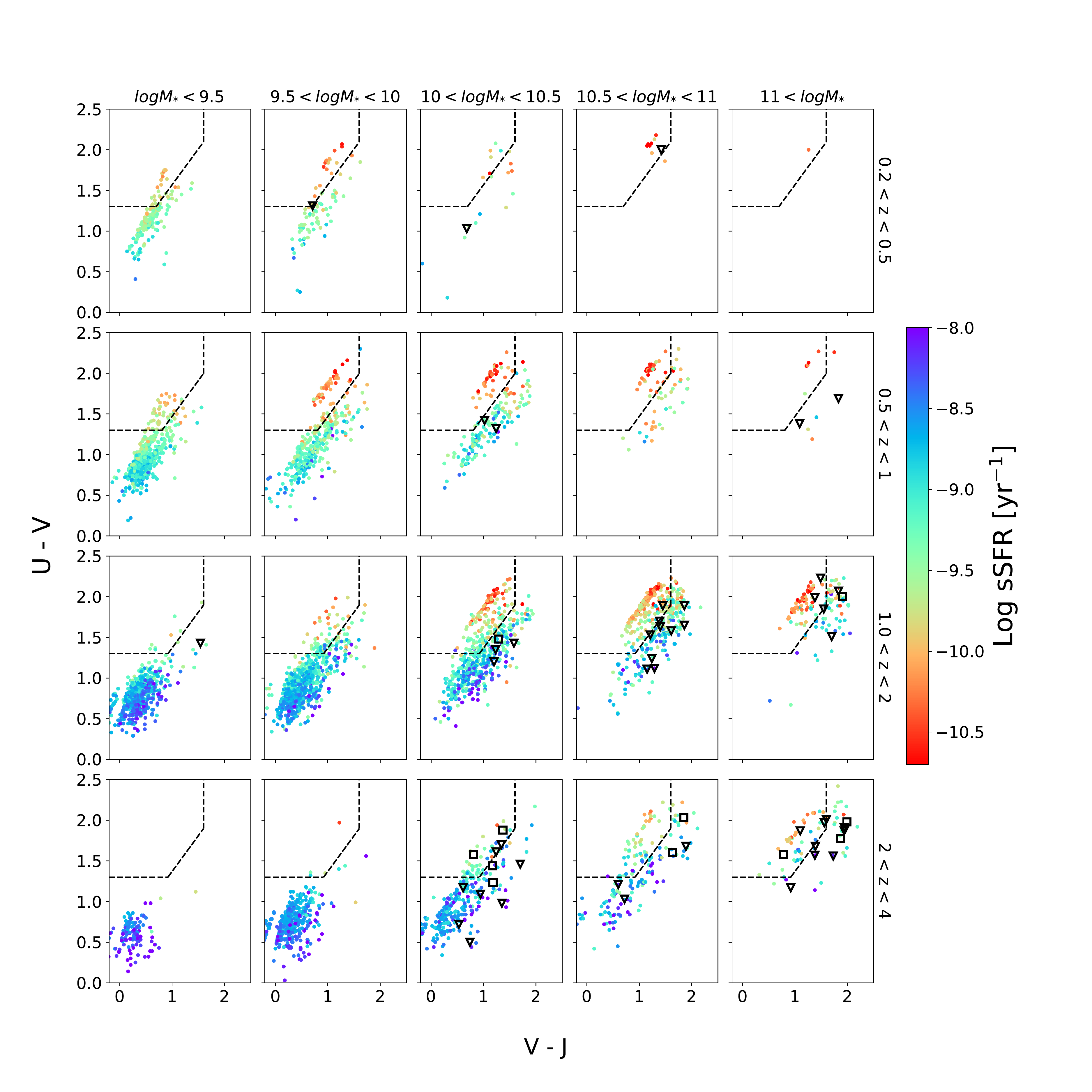}
    \caption{\textit{UVJ} color diagram of EGS CANDELS galaxies, highlighting the optical counterparts of submillimeter galaxies. The columns separate the galaxies according to their stellar mass, and the rows according to their redshift. The color gradient shows the $\log \rm sSFR$ derived from optical to min-IR data. The empty triangles are the optical counterparts of SMGs that comply with the selection criteria and the empty squares are the SMGs that do not comply with the selection criteria. The region selected by the dashed lines marks the location of quiescent galaxies.}
    \label{UVJ_allM_allz}
\end{figure*}

In order to select the star forming galaxies in the EGS field we use the $U-V$ vs $V-J$ color diagram. In Figure \ref{UVJ_allM_allz} we plot the $UVJ$ color diagram for the five stellar mass and four redshift bins considered, including all CANDELS galaxies with well-constrained parameters (as described in section \ref{sec:sample_select}). We apply the selection criteria by \citet{Williams2009} at the corresponding redshifts. At $2.5<z<3$ the galaxies with $\log (M_{\star}/\rm M_{\odot})>10.5$ start to populate the quiescent region. At lower redshifts the quiescent region is more populated by galaxies with lower stellar mass. 

\break \newpage
\section{Mann-Whitney test results}

This appendix provides the results of the different Mann-Whitney tests performed on the physical properties and morphological parameter distributions presented in this work: stellar mass (Table \ref{tab:med_massApp}), residual SFR to main sequence (Tables \ref{tab:med_deltasfr_iruv} and \ref{tab:med_deltasfr_fir}), parametric morphology (Table \ref{tab:pvalues}) and non-parametric morphology (Table \ref{tab:CAS_pvalues}).
\begin{table}
    \begin{center}
    \caption{Median stellar mass of the optically-selected star-forming galaxy (SFG) and faint submillimeter galaxy (SMG) samples. The columns are (1) redshift range; (2) median stellar mass of SFGs; (3) median stellar mass of faint SMGs; (4) probability ($p$) of the Mann-Whitney test that characterizes if the median stellar masses of faint SMGs and SFGs could correspond to a common parent distribution, with the upward arrows marking that the median mass of SMGs is statistically larger than that of SFGs; (5) 68-per cent confidence limits of $p$, taking into account redshift uncertainties.  We assign to each galaxy a new redshift under its redshift uncertainty distribution and recalculate the test 5000 times in order to produce this confidence limit. We highlight probabilities $<0.05$, which we adopt as a threshold to reject the null hypothesis of a common parent distribution for both galaxy samples. The errors of the median values of stellar mass were calculated with a bootstrap.}
    \label{tab:med_massApp}
    \begin{tabular}{ccccc}
        \hline
        $z$ & $\log M_{\star,\rm{SFGs}}$ & $\log M_{\star,\rm{SMGs}}$ & $p$ & $p$\\ 
        & [$\rm M_{\odot}$] & [$\rm M_{\odot}$] & & 68\% CL\\ \hline
        0.2--1.0 & $9.46\pm0.01$ & $10.3\pm0.6$ & $\uparrow~$\textbf{6x10$^{-5}$} &
        \textbf{2x10$^{-4}$ -- 3x10$^{-4}$} \\   
        1.0--2.0 & $9.70\pm0.01$ & $10.9\pm0.1$ & $\uparrow~$\textbf{1x10$^{-10}$} &
        \textbf{6x10$^{-13}$ -- 1x10$^{-10}$} \\ 
        2.0--3.0 & $9.86\pm0.02$ & $10.8\pm0.2$ & $\uparrow~$\textbf{3x10$^{-10}$} &
        \textbf{3x10$^{-13}$ -- 2x10$^{-8}$} \\    
        3.0--4.0 & $10.28\pm0.05$ & $10.5\pm0.1$ & $\uparrow~$\textbf{0.03} &
        \textbf{2x10$^{-6}$ -- 5x10$^{-3}$} \\ \hline   
        all $z$ & $9.68\pm 0.01$ & $10.8\pm0.1$ & $\uparrow~$\textbf{3x10$^{-26}$} &
        \textbf{2 -- 8 x10$^{-26}$} \\  
        \hline 
    \end{tabular}
    \end{center}
\end{table}
\begin{table}
    \begin{center}
    \caption{Median difference of specific star formation rate of SMGs $\rm{SFR_{IR+UV}}$, $\Delta \log \rm{sSFR_{SMG}}$, to the star-formation main sequence. The columns present: (1) redshift range; (2)  median $\Delta \log \rm sSFR$ of SMGs to main sequence; (3) probabilities ($p$) of the Mann-Whitney test for the sSFR difference to the main sequence of SFGs and SMGs, the upward arrows marking that the median values of SMGs are statistically larger than those of SFGs; (4) bootstrapped 68~per cent confidence interval of $p$, considering redshift uncertainties. We highlight probabilities $<0.05$, adopted as a threshold to reject the null hypothesis of a common parent distribution for both galaxy samples. }
    \label{tab:med_deltasfr_iruv}
    \begin{tabular}{cccc}
        \hline 
        $z$ & $\Delta \log \rm{sSFR_{SMGs}}$ & $p$ & $p$\\  & [$\rm yr^{-1}$] & & 68\% CL\\ \hline
        0.2--1.0 & $0.19\pm0.21$ &  0.08 & 0.55 -- 0.63\\
        1.0--2.0 & $0.38\pm0.13$ & $\uparrow~$ \textbf{3.8x10$^{-5}$} & \textbf{7x10$^{-6}$ -- 5x10$^{-5}$}\\
        2.0--3.0 & $0.40\pm0.04$ & $\uparrow~$ \textbf{7.6x10$^{-6}$} & \textbf{7x10$^{-10}$ -- 3x10$^{-6}$} \\
        3.0--4.0 & $0.51\pm0.14$ &  $\uparrow~$ 0.07 & \textbf{2x10$^{-6}$ -- 9x10$^{-3}$}\\ \hline
        all $z$ & $0.39\pm0.4$ & $\uparrow~$ \textbf{2.4x10$^{-11}$} & \textbf{ 2x10$^{-11}$ -- 5x10$^{-10}$} \\ 
        \hline
    \end{tabular}
    \end{center}
\end{table}
\begin{table}
    \begin{center}
    \caption{Median difference of specific star formation rate of SMGs $\rm{SFR_{FIR+UV}}$, $\Delta \log \rm{sSFR_{SMG}}$, to the star-formation main sequence. The columns present: (1) redshift range; (2) median $\Delta \log \rm sSFR$ of SMGs to main sequence; (3) probabilities ($p$) of the Mann-Whitney test for the sSFR difference to the main sequence of SFGs and SMGs, the upward arrows marking that the median values of SMGs are statistically larger than those of SFGs; (4) bootstrapped 68~per cent confidence interval of $p$, considering redshift uncertainties. We highlight probabilities $<0.05$, adopted as a threshold to reject the null hypothesis of a common parent distribution for both galaxy samples.}
    \label{tab:med_deltasfr_fir}
    \begin{tabular}{cccc}
        \hline 
        $z$ & $\Delta \log \rm{sSFR_{SMGs}}$ & $p$ & $p$\\  & [$\rm yr^{-1}$] & & 68\% CL\\ \hline
        0.2--1.0 & $-0.06\pm0.3$ & 0.9 & 0.32-0.37\\
        1.0--2.0 & $0.21\pm0.12$ & $\uparrow$\textbf{7x10$^{-3}$} & \textbf{1x10$^{-3}$ -- 3x10$^{-3}$} \\
        2.0--3.0 & $0.44\pm0.11$ & $\uparrow~$ \textbf{1x10$^{-4}$} & \textbf{9x10$^{-10}$ -- 5x10$^{-7}$} \\
        3.0--4.0 & $0.40\pm0.05$ &  $\uparrow~$ \textbf{2x10$^{-4}$} & \textbf{2x10$^{-7}$ -- 7x10$^{-5}$}\\ \hline
        all $z$ & $0.37\pm0.06$ & $\uparrow~$ \textbf{3x10$^{-8}$} & \textbf{3x10$^{-8}$ -- 1x10$^{-7}$} \\ 
        \hline
    \end{tabular}
    \end{center}
\end{table}
\begin{table*}
    \caption{Probabilities of the Mann-Whitney tests applied to the distributions of structural parameters of SMGs and SFGs with $\log (M_{\star}/\rm M_{\odot}) >10$. The top row shows the redshift bins considered in the analysis. Rows 2 and 3 give the number of SMGs and SFGs at each redshift bin. Rows 4-11 give the probabilities for the null hypothesis of identity of the medians of structural parameters to be true. We state the values for the best redshifts estimated for the SMG and SFG samples, and also for the bootstrap when we randomly assign a redshift for each galaxy within its uncertainty distribution. We highlight in bold the values that reject the null hypothesis of a common parent distribution for SMGs and SFGs taking into account best and randomized redshifts within their uncertainties, and the upward arrows mark when the median values of SMGs are statistically larger than those of SFGs.}
    \label{tab:pvalues}
    \begin{tabular}{l|c|c|c|c|c}
        \hline
        Redshift bin & all $z$ & $0.2 < z < 1.0$ & $1.0 < z < 2.0$ & $2.0<z<3.0$ & $3.0 < z < 4.0$ \\ \hline
        Num. submillimeter galaxies  & 55 & 6 & 20 & 20 & 9\\ 
        Num. star-forming galaxies & 1349 & 201 & 693 & 367 & 88 \\ \hline
        Effective radius ($R_{\rm{e,SMA}}$,best $z$) & $\uparrow~$\textbf{7x10$^{-6}$} & 0.08 & $\uparrow~$\textbf{1x10$^{-3}$} & $\uparrow~$\textbf{2x10$^{-4}$} &  0.07 \\
        Effective radius ($R_{\rm{e,SMA}}$) & \textbf{3x10$^{-7}$ -- 7 x10$^{-7}$} &  7x10$^{-3}$ -- 0.02 & \textbf{ 2x10$^{-5}$ -- 1x10$^{-3}$} & \textbf{ 7x10$^{-6}$ -- 9x10$^{-3}$} & 0.02 -- 0.2\\ \hline
        Residual effective radius ($\Delta\log R_{\rm{e,SMA}}$,best $z$) & $\uparrow~$\textbf{1x10$^{-3}$} & 0.13 & 0.14 & $\uparrow~$\textbf{4x10$^{-3}$} & 0.16 \\
        Residual effective radius ($\Delta\log R_{\rm{e,SMA}}$) & \textbf{ 1x10$^{-4}$ -- 8x10$^{-4}$} &  6x10$^{-3}$ -- 0.01 &  0.06 -- 0.1 & \textbf{ 4x10$^{-4}$ -- 0.04} & 0.06 -- 0.2 \\ \hline
        S\'ersic index ($n$,best $z$) & 0.11 & 0.13 & 0.17 & 0.29  & 0.15 \\
        S\'ersic index ($n$) & 0.1 - 0.12 & 0.08 -- 0.11 & 0.12 -- 0.23 & 0.32 -- 0.47 & 0.08 -- 0.35 \\ \hline
        Axis ratio ($q$,best $z$) & $\uparrow~$ \textbf{0.04} & 0.3  & $\uparrow~$ \textbf{0.01} & 0.3  & 0.1\\
        Axis ratio ($q$) & \textbf{0.03 -- 0.05} & 0.3 -- 0.4 & \textbf{ 7x10$^{-3}$ -- 0.02} & 0.3 -- 0.5 & 0.06-- 0.25 \\ \hline
    \end{tabular}
\end{table*}
\begin{table*}[h]
    \centering
    \caption{Probabilities of the Mann-Whitney test that measure if the medians of the SFG and SMG indices can be derived from the same parent distribution. The top row shows the redshift bins considered in the analysis. Rows 2 and 3 give the number of SMGs and SFGs at each redshift bin. Rows 4-9 give the probabilities for the null hypothesis of identity of the medians of non-parametric indices to be true for both the best-$z$ and the 68~percent confidence intervals of $p$ considering redshift uncertainties. We highlight in bold the values that reject the null hypothesis of a common parent distribution for SMGs and SFGs. Upward arrows mark the bins where the median values of SMGs are statistically larger than those of SFGs, and downward arrows where the median values of SMGs are statistically smaller than those of SFGs. } 
    \begin{tabular}{l|c|c|c|c|c}
        \hline
        Redshift  & $0.2<z<4$ & $0.2<z<1.0$ & $1.0<z<2.0$ & $2.0<z<3.0$ & $3.0<z<4.0$ \\\hline
        Num. SMGs & 54 & 6 & 20 & 19 & 9 \\
        Num. SFGs & 361 & 51 & 203 & 96 & 11 \\ \hline
        C (best $z$) & $\downarrow~$\textbf{0.03}  &  $\uparrow~$0.05 & 0.19 & $\downarrow~$\textbf{9x10$^{-4}$} & 0.06 \\
        C  & \textbf{0.03--0.046}  & 0.05--0.06  &  0.17 --0.27  &  \textbf{3x10$^{-5}$ -- 6x10$^{-4}$}  &  0.03--0.15  \\
        A (best $z$)  & 0.13  &  0.09 & 0.48 & 0.04 & 0.13 \\
        A  &  0.11--0.15  &  0.10--0.14  &  0.42--0.50  &  0.04--0.14  &  0.02--0.11  \\
        S (best $z$)  & 0.22  &  0.19 & 0.22 & 0.44 & 0.47 \\ 
        S  &  0.20--0.23  &  0.18--0.21  &  0.13--0.28  &  0.32--0.48  &  0.3--0.5  \\
        \hline
    \end{tabular}
    \label{tab:CAS_pvalues}
\end{table*}


\bsp	
\label{lastpage}

\end{document}